\documentclass[trackchanges]{aastex701}

\usepackage{array} 
\newcolumntype{T}{>{\scriptsize}c} 

\begin{document}

\title{A study of the Physical Properties and Star Formation Activity of a Large Sample of Molecular Clouds: I Distances.}

\correspondingauthor{Min Fang}
\email{mfang@pmo.ac.cn}

\author[orcid=0009-0008-7695-4747]{Juan Mei}
\affiliation{Purple Mountain Observatory, Chinese Academy of Sciences, 10 Yuanhua Road, 210023 Nanjing, China}
\affiliation{University of Science and Technology of China, Chinese Academy of Sciences, Hefei 230026, China}
\email{juanmei@pmo.ac.cn}  

\author[orcid=0000-0001-8060-1321]{Min Fang}  
\affiliation{Purple Mountain Observatory, Chinese Academy of Sciences, 10 Yuanhua Road, 210023 Nanjing, China}
\affiliation{University of Science and Technology of China, Chinese Academy of Sciences, Hefei 230026, China}
\email{mfang@pmo.ac.cn}

\author[orcid=0000-0002-6388-649X]{Miaomiao Zhang}
\affiliation{Purple Mountain Observatory, Chinese Academy of Sciences, 10 Yuanhua Road, 210023 Nanjing, China}
\affiliation{University of Science and Technology of China, Chinese Academy of Sciences, Hefei 230026, China}
\email{miaomiao@pmo.ac.cn}

\author[orcid=0000-0003-4586-7751]{Qing-Zeng Yan}
\affiliation{Purple Mountain Observatory, Chinese Academy of Sciences, 10 Yuanhua Road, 210023 Nanjing, China}
\email{qzyan@pmo.ac.cn}

\author[orcid=0000-0003-0804-9055]{Lixia Yuan}
\affiliation{Purple Mountain Observatory, Chinese Academy of Sciences, 10 Yuanhua Road, 210023 Nanjing, China}
\email{lxyuan@pmo.ac.cn}

\author[orcid=0000-0003-3151-8964]{Xuepeng Chen}
\affiliation{Purple Mountain Observatory, Chinese Academy of Sciences, 10 Yuanhua Road, 210023 Nanjing, China}
\affiliation{University of Science and Technology of China, Chinese Academy of Sciences, Hefei 230026, China}
\email{xpchen@pmo.ac.cn}

\author[orcid=0000-0003-0849-0692]{Zhiwei Chen}
\affiliation{Center for Astronomy and Space Sciences, China Three Gorges University, 8 University Road, 443002 Yichang, China}
\affiliation{College of Science, China Three Gorges University, 8 University Road, 443002 Yichang, China}
\email{zwchen@pmo.ac.cn}

\author[orcid=0000-0002-0197-470X]{Yang Su}
\affiliation{Purple Mountain Observatory, Chinese Academy of Sciences, 10 Yuanhua Road, 210023 Nanjing, China}
\affiliation{University of Science and Technology of China, Chinese Academy of Sciences, Hefei 230026, China}
\email{yangsu@pmo.ac.cn}

\author[orcid=0009-0002-2379-4395]{Shiyu Zhang}
\affiliation{Purple Mountain Observatory, Chinese Academy of Sciences, 10 Yuanhua Road, 210023 Nanjing, China}
\affiliation{University of Science and Technology of China, Chinese Academy of Sciences, Hefei 230026, China}
\email{syzhang@pmo.ac.cn}

\author[orcid=0000-0002-5920-031X]{Zhibo Jiang}
\affiliation{Purple Mountain Observatory, Chinese Academy of Sciences, 10 Yuanhua Road, 210023 Nanjing, China}
\email{zbjiang@pmo.ac.cn}

\author[orcid=0000-0001-7768-7320]{Ji Yang}
\affiliation{Purple Mountain Observatory, Chinese Academy of Sciences, 10 Yuanhua Road, 210023 Nanjing, China}
\email{jiyang@pmo.ac.cn}


\begin{abstract}


Accurate distances to molecular clouds are crucial for determining their physical properties, understanding star formation, and tracing Galactic spiral structure. A number of 103,517 molecular clouds has been identified by the DBSCAN algorithm in the MWISP Phase I CO survey ($l=9.75\degr$–$229.75\degr$, $|b| \leq 5.25\degr$), most of which lack reliable distances. In this work, we propose three independent methods, all of which match the molecular cloud's velocity-integrated intensity maps of $^{12}$CO lines from the MWISP with the three-dimensional dust extinction maps derived from Gaia, Pan-STARRS 1, and 2MASS, to determine molecular cloud distances. We present a catalog of 1,573 molecular clouds with robust distances ranging from $\sim$150 pc to $\sim$3000 pc, 90 percent of which are measured for the first time, with typical statistical and systematic uncertainties of $\sim$20$\%$ and $\sim$10$\%$, respectively. We also derive their physical properties, such as their mass and sizes. This publicly available catalog of molecular clouds with distances provides a foundation for testing molecular cloud scaling relations and probing how cloud conditions influence star formation across diverse Galactic environments.

\end{abstract}


\section{Introduction}
Molecular clouds (MCs) are the primary birthplaces of stars in Galaxy \citep{Blitz+1999}, and their formation and evolution are key processes in the galactic life cycle. A statistical understanding of their physical properties (e.g., mass, size, kinematics) is essential to establish universal scaling relations, probe environmental dependencies, and constrain the physics linking cloud conditions to stellar output \citep{McKee+2007, Kennicutt+2012, HeyerDame+2015}. However, the ability to perform such quantitative, comparative studies hinges entirely on a basic parameter: distance. Accurate distances are the fundamental scaling key for deriving intrinsic cloud properties, placing clouds within the Galactic context to study large-scale structure \citep{Dame+2001, Xu+2018}, and advancing beyond a qualitative picture of star formation. Consequently, constructing a large, unbiased sample of molecular clouds with reliable distances is the critical yet challenging first step for any investigation into their physics and star-forming potential.

As the most abundant observable molecule after H$_2$, the $^{12}$CO line serves as the primary tracer for molecular clouds. The advent of large-scale CO surveys has enabled the construction of comprehensive molecular cloud catalogs, with distances predominantly derived from kinematic methods. For instance, \citet{Rice+2016} identified 1,064 high-mass clouds using dendrograms applied to the CfA–Chile survey \citep{Dame+2001}, while \citet{Miville-Deschenes+2017} applied a different hierarchical clustering algorithm to the same data, reporting a much larger sample of 8,107 clouds. In these studies, distances were calculated from line-of-sight velocities ($V_{\mathrm{LSR}}$) using Galactic rotation models \citep{Brand+1993, Reid+2014}, an approach subject to the well-known kinematic distance ambiguity (KDA) and significant systematic uncertainties, particularly in the inner Galaxy. In parallel, cloud distances can also be estimated through association with objects of known distance, such as OB associations, young open clusters, H{\sc ii} regions, young stellar objects (YSOs), and masers \citep{Xu+2006, Russeil+2007, Reid+2019, Cantat-Gaudin+2021, Marton+2022, Zhang+2023}. While these association-based methods can provide more precise distances where applicable, their use is largely restricted to well-known molecular clouds and active star-forming regions. 

The release of \textit{Gaia} astrometric data provides precise stellar parallaxes, making it possible to estimate molecular cloud distances by locating extinction break points along the line of sight toward molecular clouds. Combined with \textit{Gaia} data, several studies have constructed new three-dimensional (3D) extinction maps and estimated extinction and distances for millions of stars \citep{Green+2019, ChenBQ+2019, Lallement+2019, Guo+2021, Sun+2021a, Wang+2025a}. The extinction method relies on accurately estimating the distance and extinction of numerous stars. This method is implemented in two ways: one involves modeling the dust extinction profiles along different lines of sight (LOS) to identify dust clouds from 3D extinction map \citep{ChenBQ+2020a, Guo+2022, Xie+2024, Wang+2025b}, while the other involves identifying an extinction break point that is caused by molecular clouds \citep{Yan+2019, Zucker+2019, Zucker+2020, ChenBQ+2020b, Yan+2021, Sun+2021b, Sun+2024, Mei+2024}. These studies have yielded distances for thousands of dust clouds, yet the number of molecular clouds with reliably determined distances remains limited. Thus, constructing a comprehensive catalog of molecular clouds with precise distances is imperative. 

The first data release (DR1) of the Milky Way Imaging Scroll Painting (MWISP\footnote{\url{http://www.radioast.nsdc.cn/mwisp.php}}) survey within the Galactic longitude range $l=9.75\degr-229.75\degr$ and latitude $|b| \leq 5.25\degr$ is uniquely positioned to address this need. By covering the entire northern Galactic plane with its high sensitivity and angular resolution, MWISP provides an unprecedented, uniform census of molecular gas. A total of 103,517 molecular clouds has been identified within the MWISP DR1 data, and distances for 234 of them have been measured by \cite{Yan+2021} using the background-eliminated extinction‑parallax method. However, precise distances remain unavailable for the majority of these clouds. In this work, we explore three independent methods that combine $^{12}$CO integrated intensity maps from MWISP with the 3D extinction maps of \citet{Green+2019} to determine molecular cloud distances. Applying these to the MWISP DR1 data, we present a comprehensive catalog of 1,573 molecular clouds with distances ranging from $\sim$150 to 3000 pc, the majority of which are measured for the first time.

The paper is structured as follows. In Section 2, we describe the data. In Section 3, we introduce our methods for estimating the distances of molecular clouds. We present our results and discussions in Section 4. We summarize the conclusions in Section 5.

\section{Data}
\subsection{CO data and cloud samples}
The Milky Way Imaging Scroll Painting (MWISP) project is a multi-line survey observing $^{12}$CO, $^{13}$CO, and C$^{18}$O ($J=1-$0) simultaneously using the Purple Mountain Observatory (PMO) 13.7\,m telescope \citep[see details in][]{Su+2019}. The survey's Phase I data covers the northern Galactic plane ($9.75\degr \leq l \leq 229.75\degr$, $|b| \leq 5.25\degr$) within a velocity range of $-200 \leq V_{\mathrm{LSR}} \leq 300$\,km\,s$^{-1}$. In this study, we utilize the $^{12}$CO emission as the primary tracer, as it exhibits stronger and more extended structures compared to the other isotopologues. The $^{12}$CO data have a grid spacing of $30\arcsec$, a velocity resolution of $0.16\,\mathrm{km\,s^{-1}}$, and a typical rms noise of $\sim 0.5$\,K. At the frequency of the $^{12}$CO transition ($\sim 115$\,GHz), the half-power beamwidth (HPBW) of the telescope is approximately $50\arcsec$.

Based on the released MWISP $^{12}$CO data cubes, \citet{YangMWISP+2026} established a catalog of 103,517 molecular cloud samples identified using the DBSCAN algorithm \citep{Ester+1996, Yan+2020}. Our goal is to construct a catalog of molecular clouds with reliably estimated distances based on this sample. However, distance estimation for small clouds is often hindered by large statistical uncertainties arising from insufficient stellar counts and the limited resolution of extinction maps. Therefore, we impose an area threshold and select a subset of 10,929 molecular clouds with angular areas $> 0.01$\,deg$^{2}$ for our analysis.

\subsection{\textit{Gaia} StarHorse EDR3}
We adopt the StarHorse EDR3 catalog \citep{Anders+2022} to obtain the necessary stellar parameters. This catalog provides heliocentric distances $d$, $V$-band extinctions A$_{\rm V}$ ($\lambda = 542$\,nm), effective temperatures, masses, and ages for approximately 362 million stars. These parameters are derived using the StarHorse code \citep{Queiroz+2018} by combining high-precision astrometry from \textit{Gaia} EDR3 \citep{Gaia3+2021} with multi-band photometry from Pan-STARRS1, SkyMapper, 2MASS, and AllWISE.

Benefiting from the high quality of the \textit{Gaia} EDR3 data and updated stellar-density priors, the catalog achieves a typical distance uncertainty of 3\% for stars at $G = 14$ and 15\% at $G = 17$ \citep{Anders+2022}. To ensure the robustness of our analysis, we apply relatively strict selection criteria to the sample. We select stars with relative distance uncertainties better than 10\% ($\Delta d/d \leq 0.1$). Additionally, we require the quality flags \texttt{StarHorse.FlagOut} $= 0$ and \texttt{StarHorse.fidelity} $= 1$ to exclude unreliable sources.

\section{Method} 
To determine molecular cloud distances, we developed three methods that combine the $^{12}$CO integrated intensity map from the MWISP survey with the 3D extinction map from \citet{Green+2019}. The MWISP G211.613+02.405+007.49 cloud is used as an illustrative example to demonstrate the application of these methods. These approaches are designed to provide mutual cross-validation, thereby enhancing the reliability of the distance estimates and extending distance measurements to a larger sample of molecular clouds.

\subsection{Statistical analysis of extinction–distance distribution differences for cloud distances}
\label{sect:method1}
Due to dust absorption, stars located behind a molecular cloud exhibit higher extinction compared to foreground stars. Our first method determines the cloud distance by comparing the extinction-distance profiles of stars inside the cloud boundary (on-cloud) with those in a local reference region (boundary-adjacent off-cloud). We simulate the presence of the cloud by adding random Gaussian extinction to the background stars in the reference region at various assumed distances. The distance of the cloud is then defined as the value where the adjusted off-cloud distribution shows the minimum statistical difference from the observed on-cloud data.

\begin{figure*}[!htbp]
    \centering
	\includegraphics[width=0.95\linewidth]{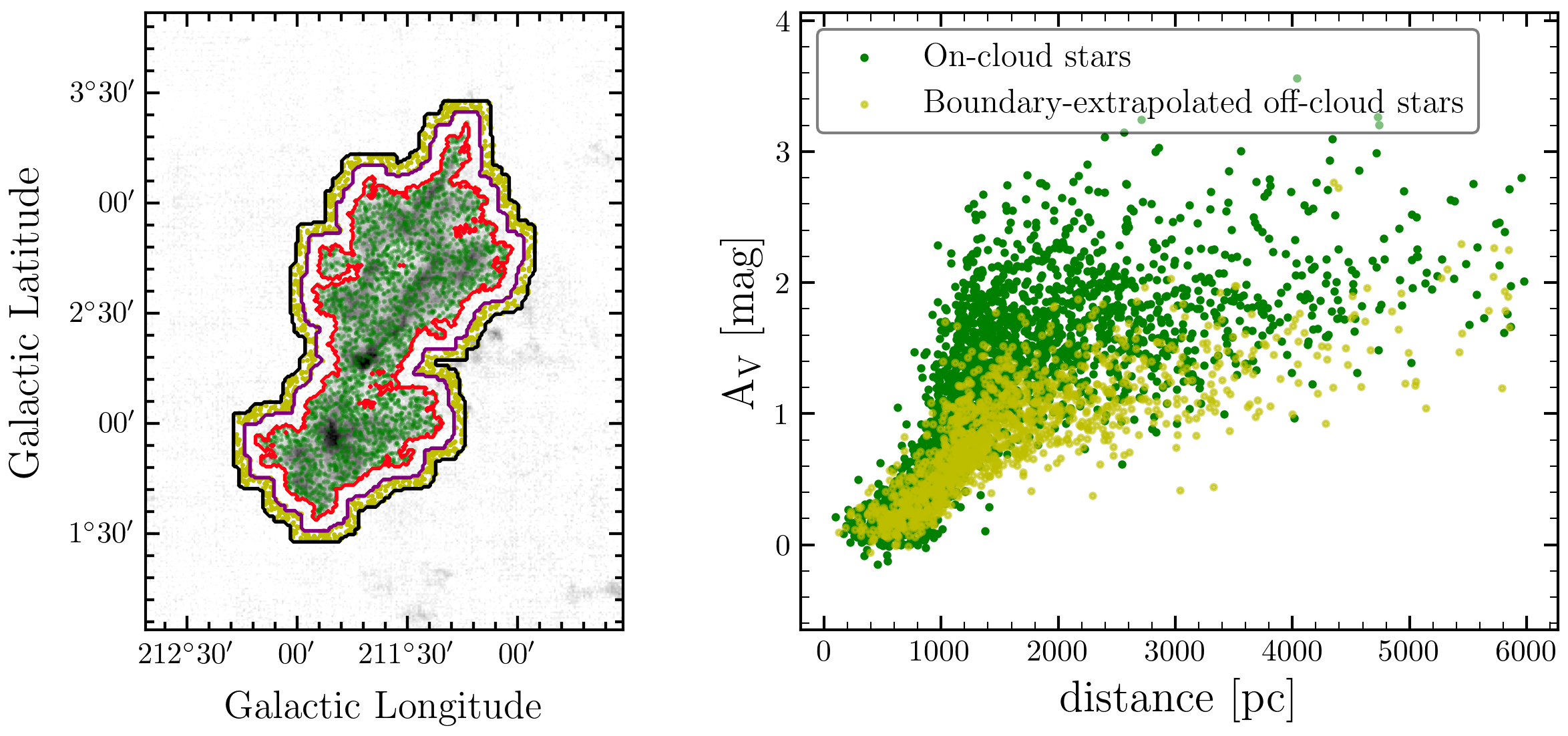}
    \caption{The left panel shows selected on-cloud stars (green) and boundary-adjacent off-cloud stars (yellow) for MWISP G211.613+02.405+007.49. The red, purple, and black contours mark the cloud boundary, and its first and second extrapolated boundaries. The right panel displays the distributions of $\mathrm {A_V}$ vs. distance for the corresponding stellar populations.}
    \label{selectstars_inall_outdouble}
\end{figure*}

To implement this approach, we first define two stellar samples for each molecular cloud using the StarHorse catalog. The ``on-cloud'' sample comprises stars projected within the cloud boundary defined by the MWISP data (green points in Figure~\ref{selectstars_inall_outdouble}, left). The ``reference'' (or boundary-adjacent off-cloud) sample is selected from a surrounding annular region (yellow points). The width of this annulus scales with the cloud area, typically extending 6 pixels ($\sim 3\arcmin$) outward from the original cloud mask.

As illustrated in the right panel of Figure~\ref{selectstars_inall_outdouble}, the on-cloud stars exhibit higher extinction compared to the reference stars due to the presence of the molecular cloud. 
Our method rests on the assumption that this excess extinction follows a Gaussian distribution $N(\mu, \sigma)$, where $\mu$ and $\sigma$ are the mean and standard deviation of the extinction excess, respectively, and affects only stars located behind the cloud (at distance $d > D$).
To estimate the cloud distance $D$, we simulate the cloud's screening effect on the reference sample. For a given set of trial parameters ($D, \mu, \sigma$), we add random extinction values drawn from $N(\mu, \sigma)$ to all reference stars located beyond the trial distance $D$. We then compare the cumulative distribution function (CDF) of this ``adjusted'' reference sample with that of the observed on-cloud sample using the Kolmogorov-Smirnov (KS) test \citep{Chakravarti+1967}. The optimal cloud distance is identified by maximizing the KS $p$-value and minimizing the KS statistic.

\begin{figure*}[!htbp]
    \centering
    \begin{minipage}{0.48\textwidth}
        \centering
        \includegraphics[width=\textwidth]{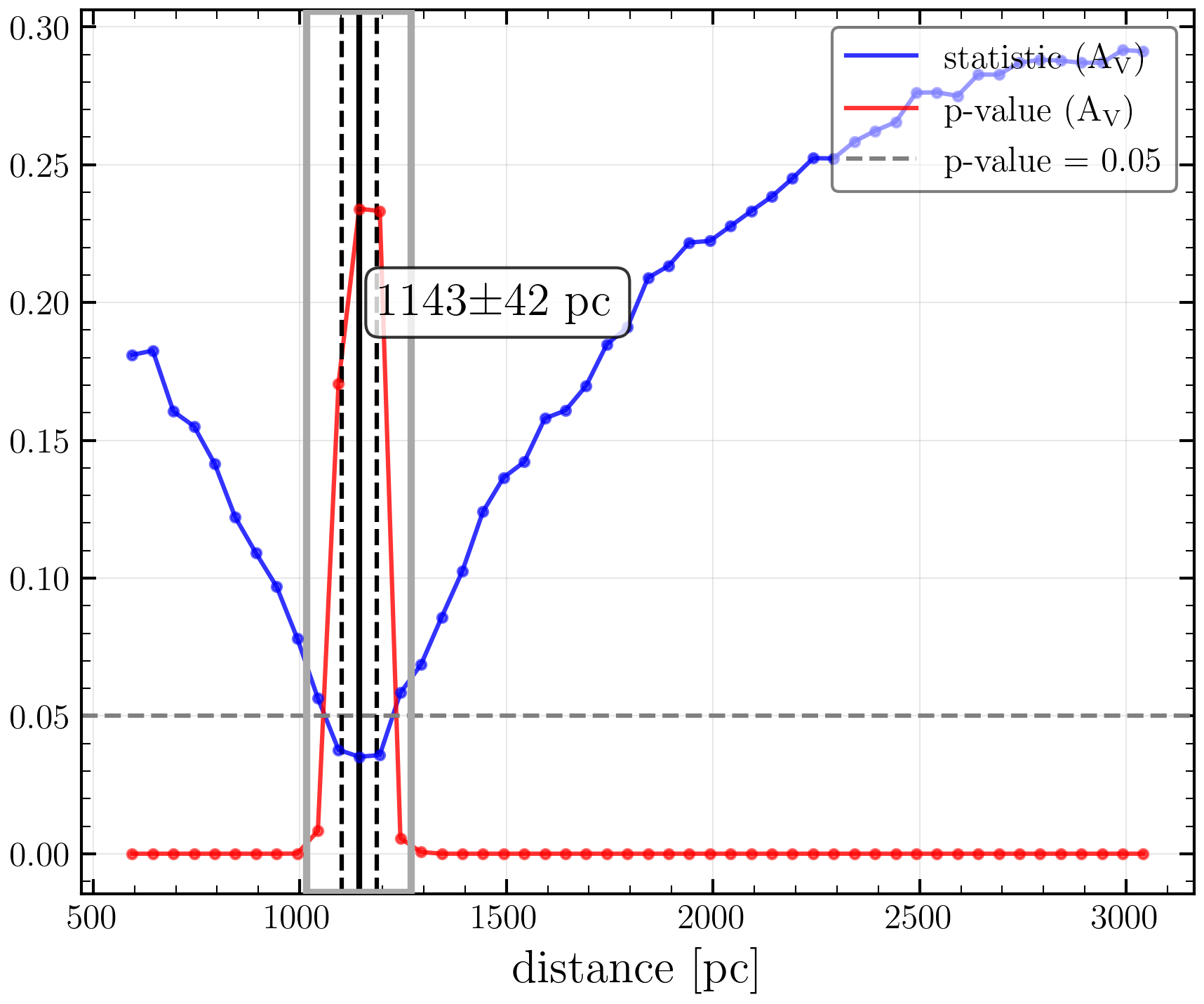}
    \end{minipage}
    \hfill
    \begin{minipage}{0.48\textwidth}
        \centering
        \includegraphics[width=\textwidth]{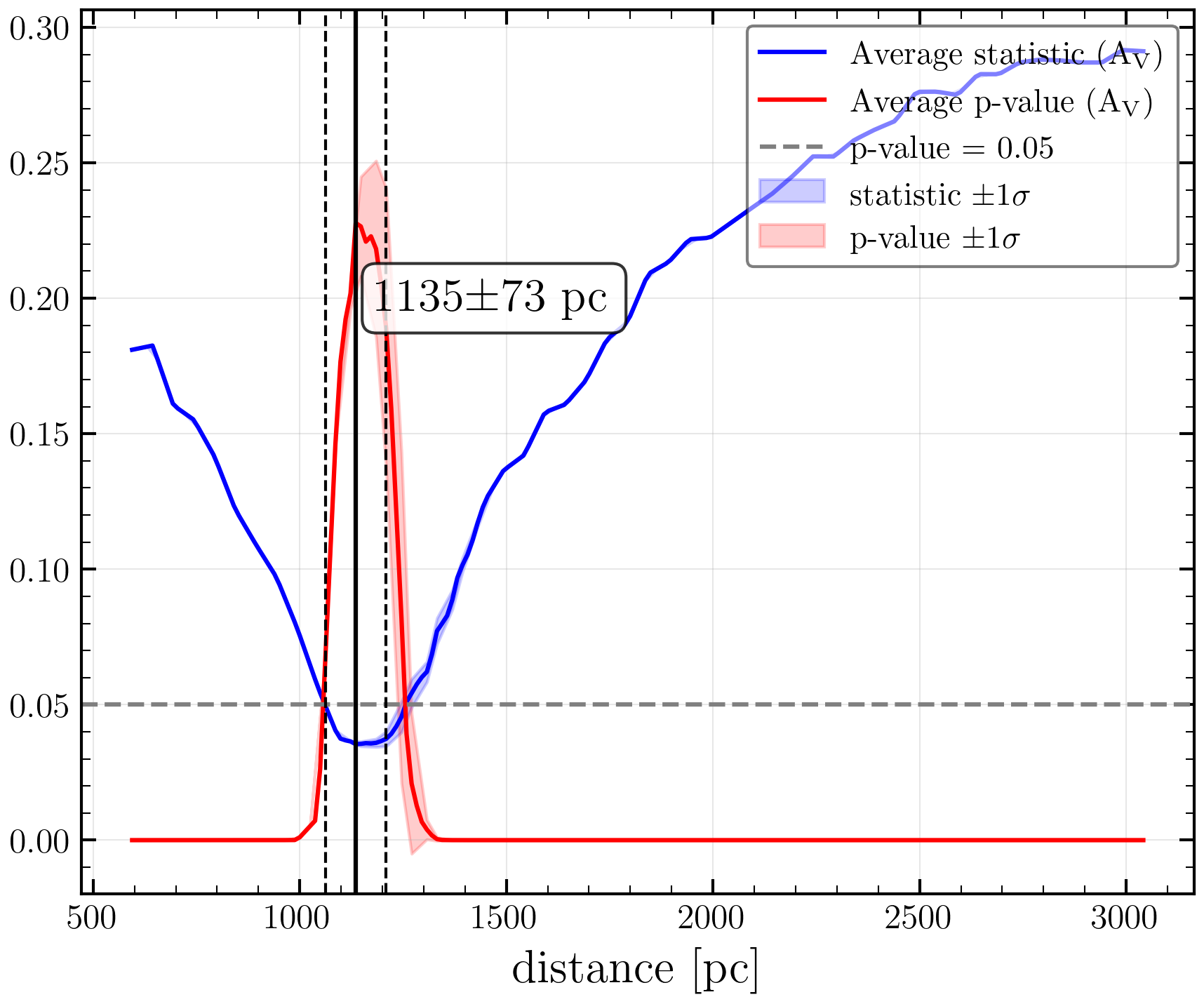}
    \end{minipage}  
    \caption{Left panel: KS test statistic (blue) and corresponding p-value (red) as functions of distance in the initial optimization. The convergence of minimum KS statistic and maximum p-value identifies a distance (black) with uncertainty, corresponding to a 3$\sigma$ confidence region (gray box) used to constrain subsequent parameter refinement. Right panel: Ensemble analysis of 100 independent Monte Carlo repetitions performed within the constrained parameter range. Average KS statistic (blue curve) and p-value (red curve) profiles are shown with 1$\sigma$ uncertainty (shaded). The final distance is indicated by the solid black vertical line, with its uncertainty range marked by the dashed black vertical lines.
    }
    \label{gridsearch_bestdistance}
\end{figure*}

We solve for the best-fit parameters in two stages. First, we perform a coarse grid search to obtain an initial estimate $(D_0, \mu_0, \sigma_0)$. We adopt uniform priors for the free parameters:
\begin{equation}
\left\{
\begin{array}{l}
    D \sim U [\mathrm{d}_{\mathrm{min}}, \mathrm{d}_{\mathrm{max}}], \\
    \mu \sim U [0.1, 2], \\
    \sigma \sim U [0.2, 2],
\end{array}
\right.
\end{equation}
where $d_{\mathrm{min}}$ and $d_{\mathrm{max}}$ denote the minimum and maximum distances of the stars in the field. This step identifies the parameter space where the KS $p$-value is maximized (Figure~\ref{gridsearch_bestdistance}, left).

Second, to refine the distance estimate and quantify its uncertainty, we execute an iterative optimization consisting of 100 independent Monte Carlo runs. In each iteration, the search ranges are dynamically centered on the optimal values from the previous step and constrained within their $3\sigma$ confidence intervals. The final molecular cloud distance is defined as the peak of the average $p$-value curve derived from these repetitions (Figure~\ref{gridsearch_bestdistance}, right). If the algorithm fails to converge to a stable solution after 10 consecutive attempts, the distance is marked as undetermined.

\begin{figure*}[!htbp]
    \centering
	\includegraphics[width=0.85\linewidth]{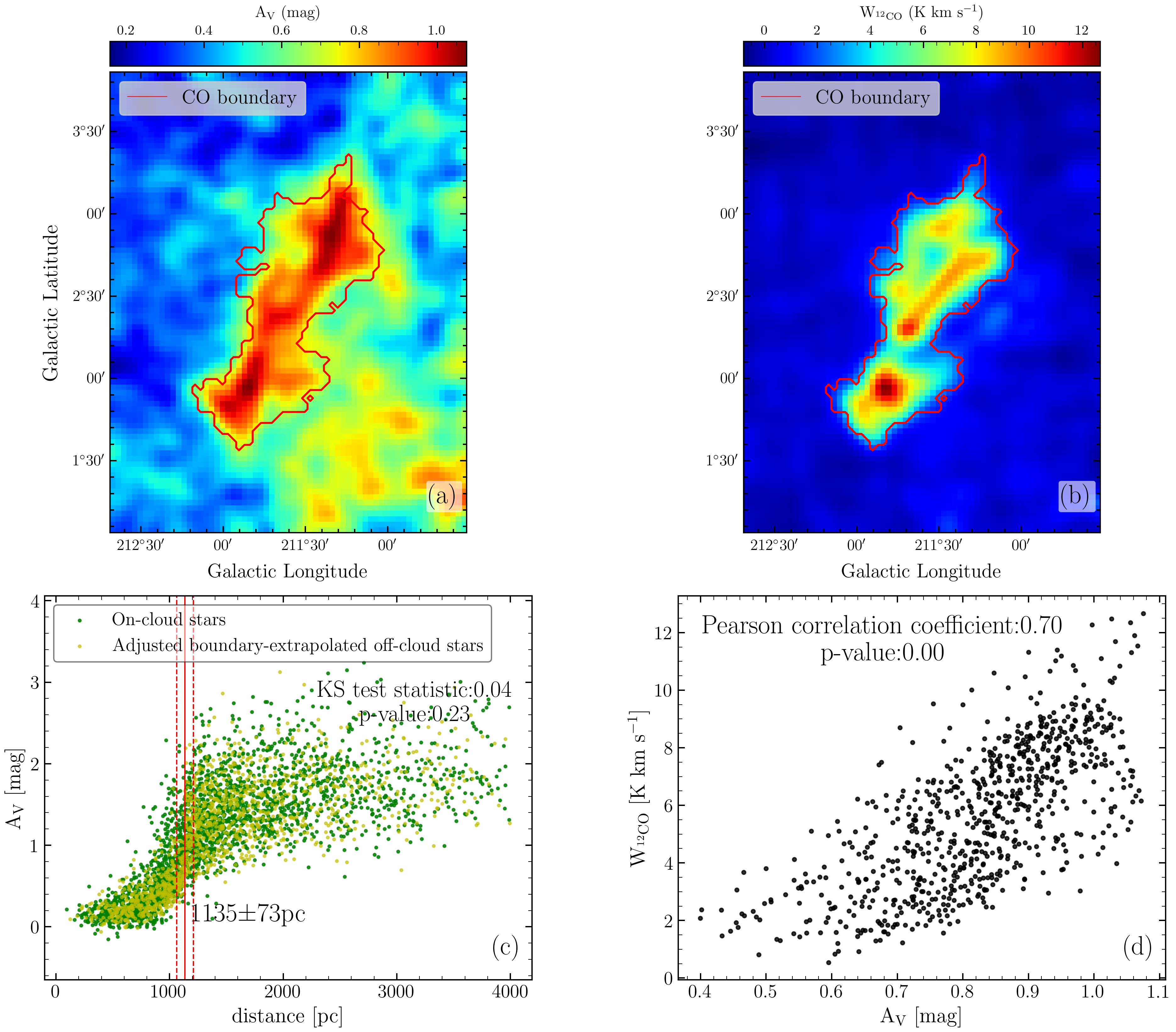}
    \caption{Distance of MWISP G211.613+02.405+007.49 based on statistical analysis of extinction-distance distributions. (a) Extinction map corresponding to $\pm 20\%$ of the measured cloud distance. (b) The $^{12}$CO integrated intensity map of the cloud. The maps in (a) and (b) are presented at a 2$^{\prime}$ resolution, with the red contour tracing the cloud boundary identified from the $^{12}$CO emission. (c) The extinction-distance distributions for on-cloud (green points) and boundary-adjacent off-cloud (yellow points) stars. The red solid line indicates the estimated distance and the red dashed lines depict the uncertainty of the cloud distance. (d) Pearson correlation between the dust extinction (A$_{\rm V}$) and the $^{12}$CO integrated intensity (W$_{\mathrm{CO}}$) at the determined distance.
    }
    \label{distance_kstest}
\end{figure*}

Finally, we validate our distance estimates by comparing the molecular gas morphology with the 3D dust extinction map from \cite{Green+2019}. We examine the spatial consistency between the $^{12}$CO integrated intensity (W$_{\mathrm{CO}}$) and the $V$-band extinction (A$_{\rm V}$) integrated within $\pm 20\%$ of the estimated distance. Both maps are regridded to a common resolution of $2\arcmin$ to facilitate comparison (Figure~\ref{distance_kstest}). A distance estimate is considered reliable only if it satisfies two criteria: (1) a visual morphological match between the CO and dust structures, and (2) a positive pixel-by-pixel Pearson correlation coefficient between W$_{\mathrm{CO}}$ and A$_{\rm V}$.

\subsection{Morphological matching of dust-cloud decomposition for distance determination}
\label{sect:method2}

The first method described in Sect.~\ref{sect:method1} is constrained by the precision of \textit{Gaia} parallaxes and the number of available stars. It becomes unreliable for very nearby clouds due to a lack of foreground stars, and for distant clouds due to large parallax errors and insufficient background stars. To address these limitations, we employ a second, independent method using the 3D dust extinction map from \citet{Green+2019}. This dataset provides extinctions and distances for 799 million stars by combining \textit{Gaia} astrometry with photometry from Pan-STARRS 1 and 2MASS. By incorporating deep photometric data, this map allows us to trace dust structures effectively even at larger distances.

\begin{figure*}[!htbp]
    \centering
    \begin{minipage}{0.45\textwidth}
        \centering
        \includegraphics[width=\textwidth]{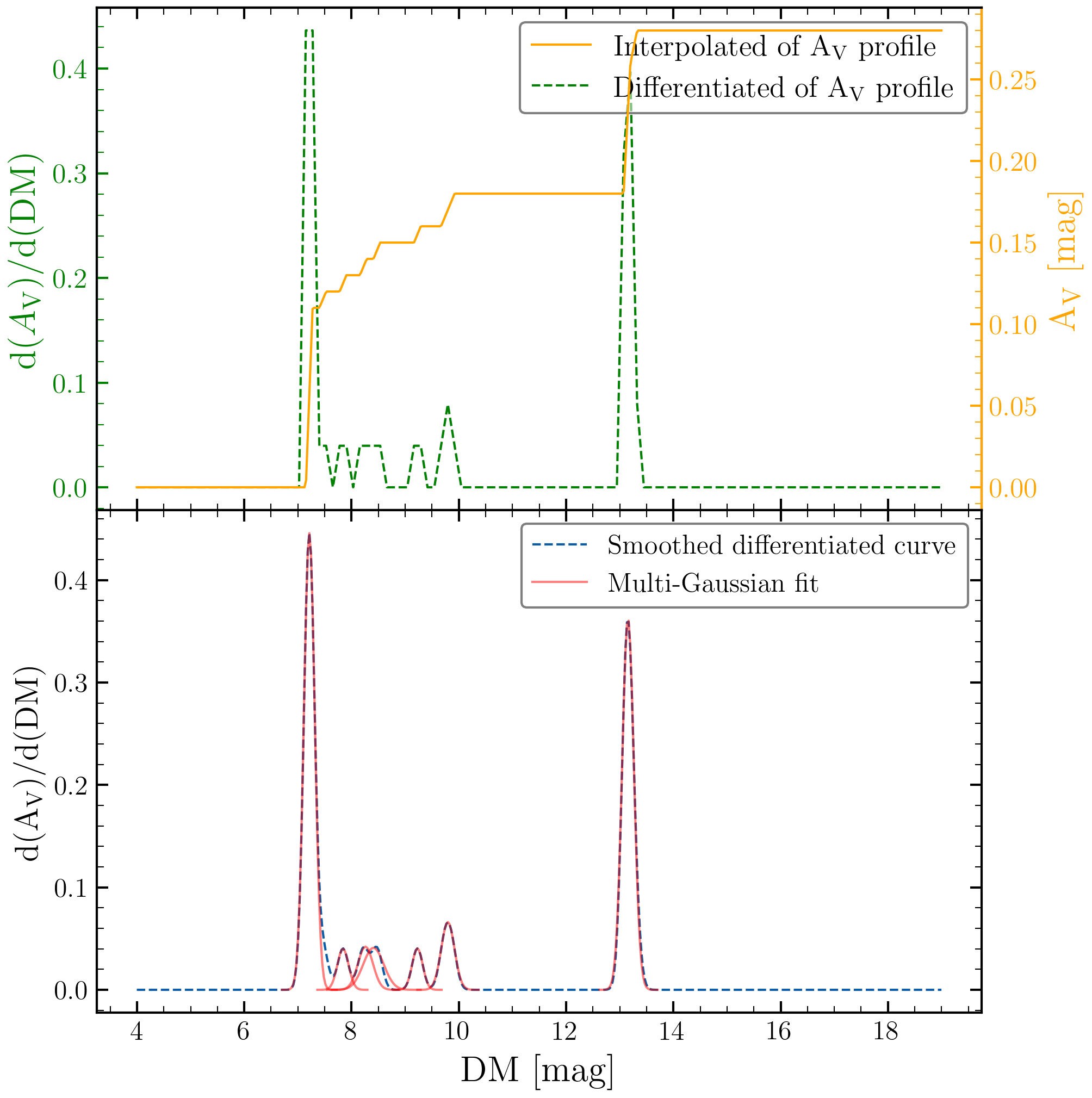}
    \end{minipage}  
    \hfill
    \begin{minipage}{0.5\textwidth}
        \centering
        \includegraphics[width=\textwidth]{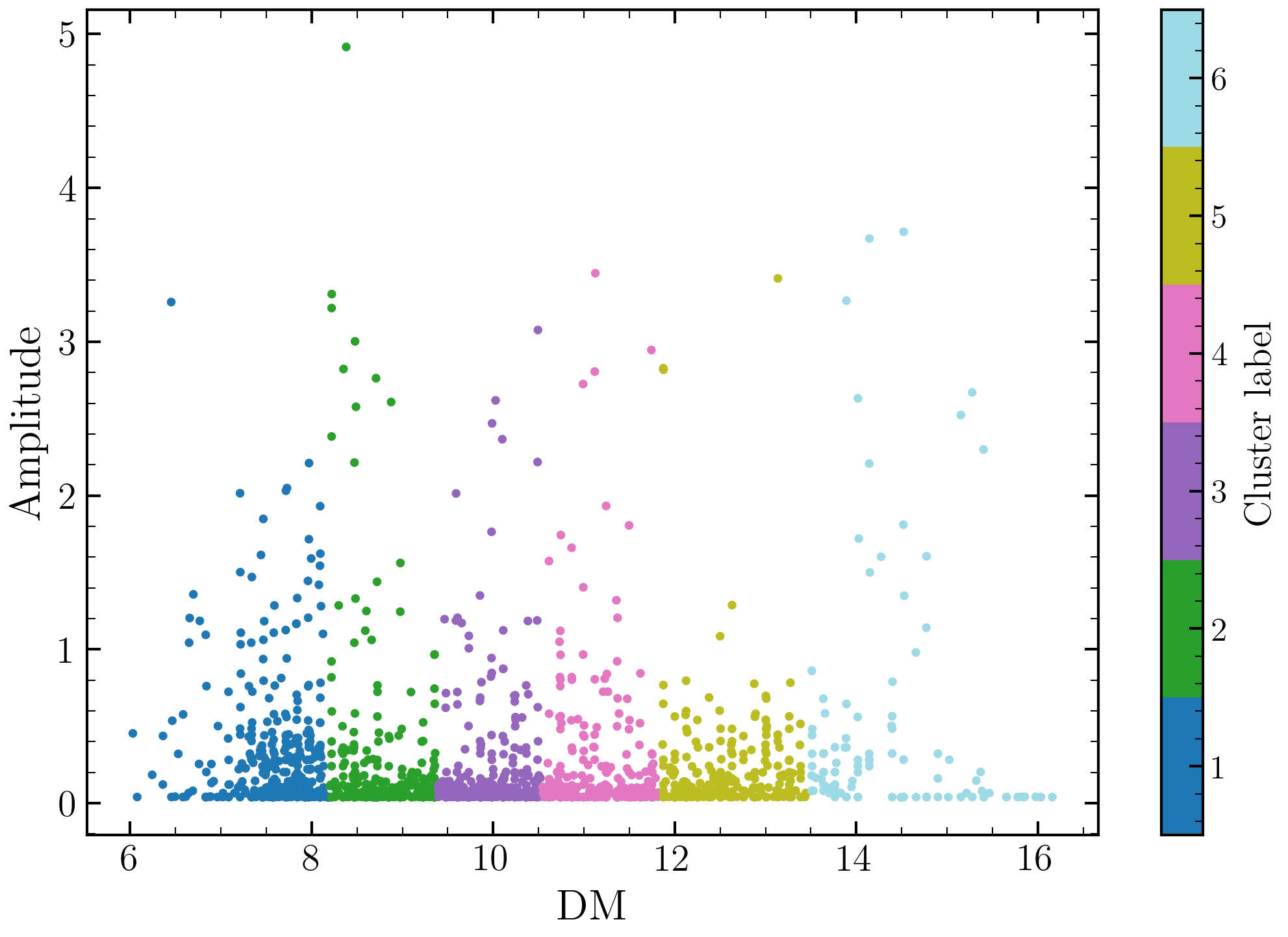}
    \end{minipage}  
    \caption{Left panel: The extinction–distance modulus relation along the example sightline. The extinction profile is interpolated (orange), differentiated (green), smoothed (blue), and fitted with a multi-Gaussian model (red). Right panel: K-means clustering results. Points represent the fitted Gaussian components, colored by cluster label.}
    \label{kmeans_gaussianfit_cluster}
\end{figure*}

Our approach determines the cloud distance by decomposing the dust cloud along the line of sight. For each molecular cloud, we examine the extinction profiles of sightlines projected within the cloud boundary. As illustrated in the left panel of Figure \ref{kmeans_gaussianfit_cluster}, we analyze the relation between extinction and distance modulus (orange curve). A sharp rise in extinction indicates a dust cloud, which corresponds to a peak in the derivative of the profile (green curve). To isolate these features, we fit the derivative with a multi-Gaussian function and extract individual Gaussian components (red curve).

We then group these components to identify coherent structures. We apply the K-means clustering algorithm to the full set of fitted Gaussian components from all sightlines, as shown in the right panel of Figure \ref{kmeans_gaussianfit_cluster}. The optimal number of clusters, $K$, is determined using the ``elbow method,'' which identifies the inflection point where the within-cluster sum of squared errors (SSE) stops decreasing significantly. The centers of these clusters represent the locations of discrete dust layers and serve as the candidate distances for the molecular cloud.

\begin{figure*}[!htbp]
    \centering
    \begin{minipage}{0.49\textwidth}
        \centering
        \includegraphics[width=\textwidth]{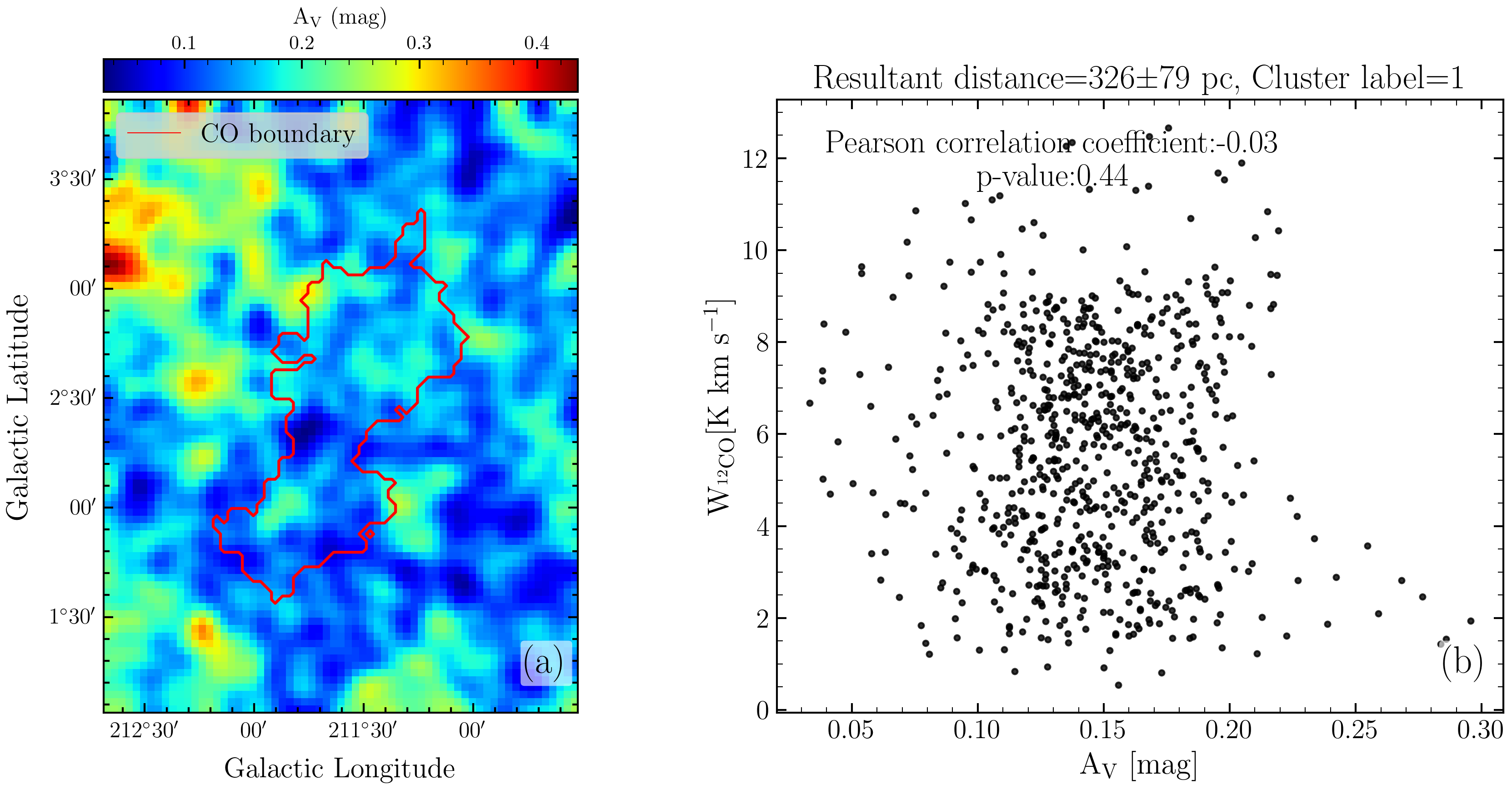}
    \end{minipage}  
    \hfill
    \begin{minipage}{0.49\textwidth}
        \centering
        \includegraphics[width=\textwidth]{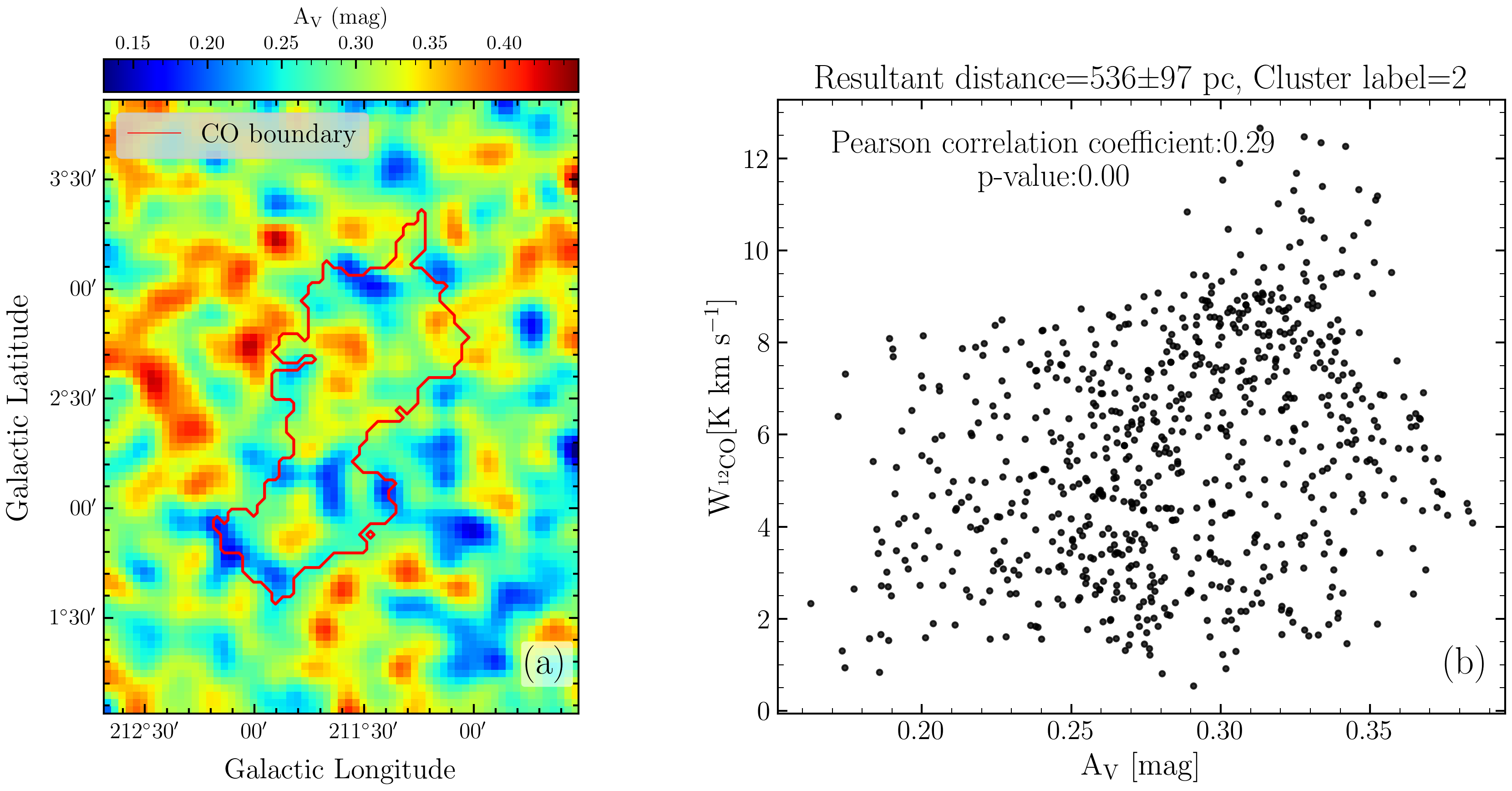}
    \end{minipage}  
    \hfill
    \begin{minipage}{0.49\textwidth}
        \centering
        \setlength{\fboxrule}{1pt} 
        \setlength{\fboxsep}{1.0 pt}  
        \fcolorbox{red}{white}{\includegraphics[width=\textwidth]{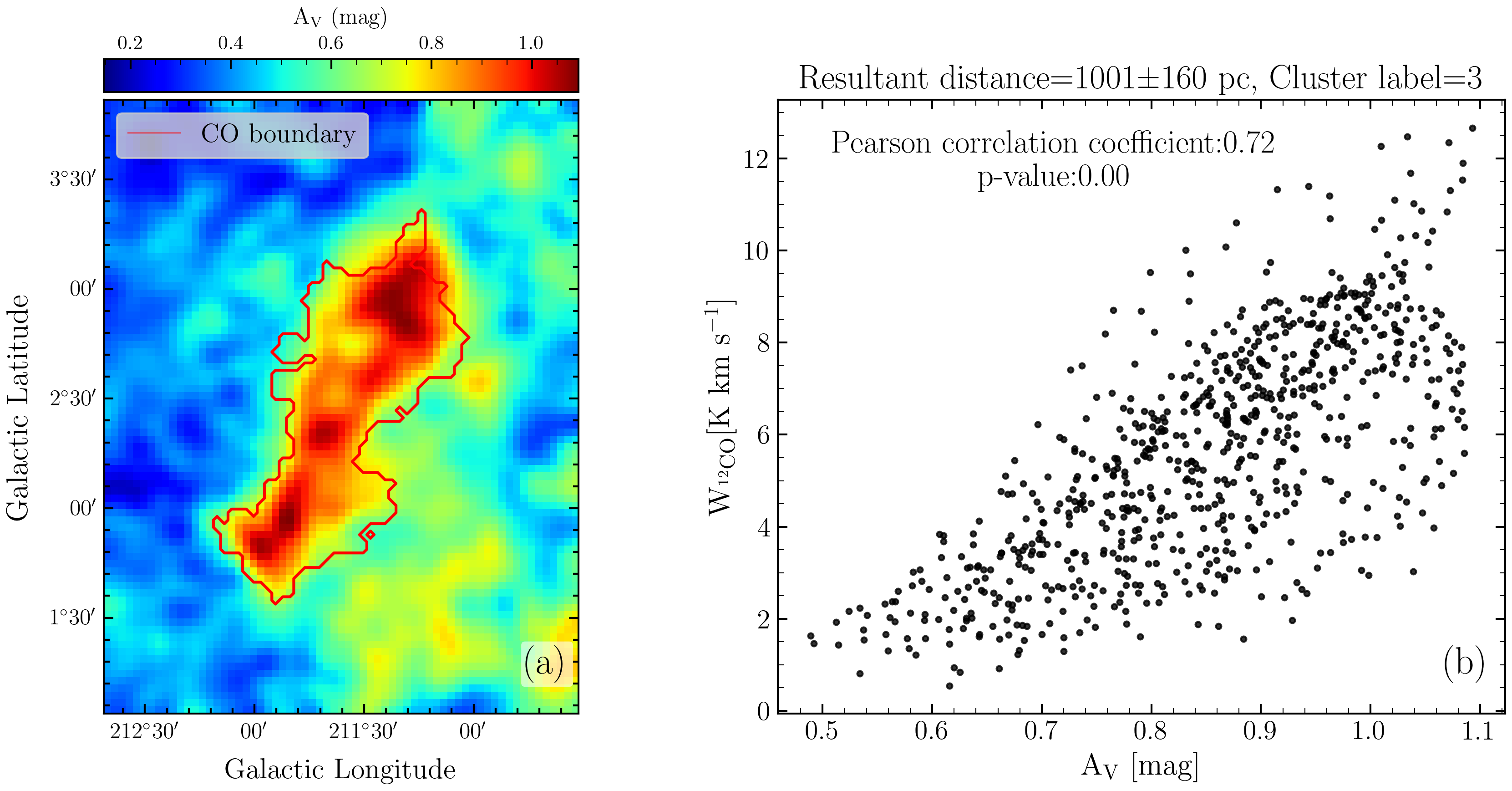}}
    \end{minipage}  
    \hfill
    \begin{minipage}{0.49\textwidth}
        \centering
        \includegraphics[width=\textwidth]{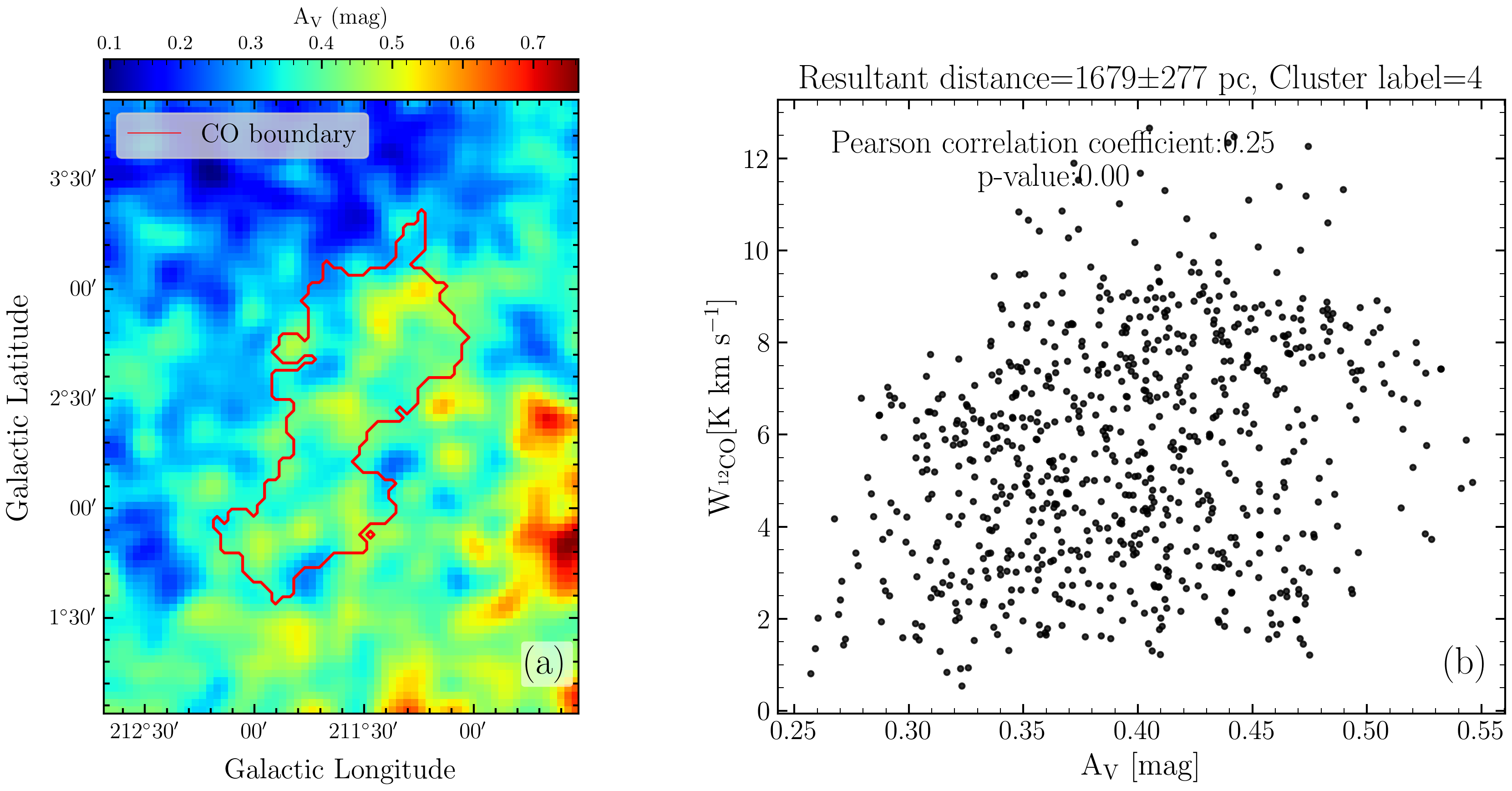}
    \end{minipage}  
    \hfill
    \begin{minipage}{0.49\textwidth}
        \centering
        \includegraphics[width=\textwidth]{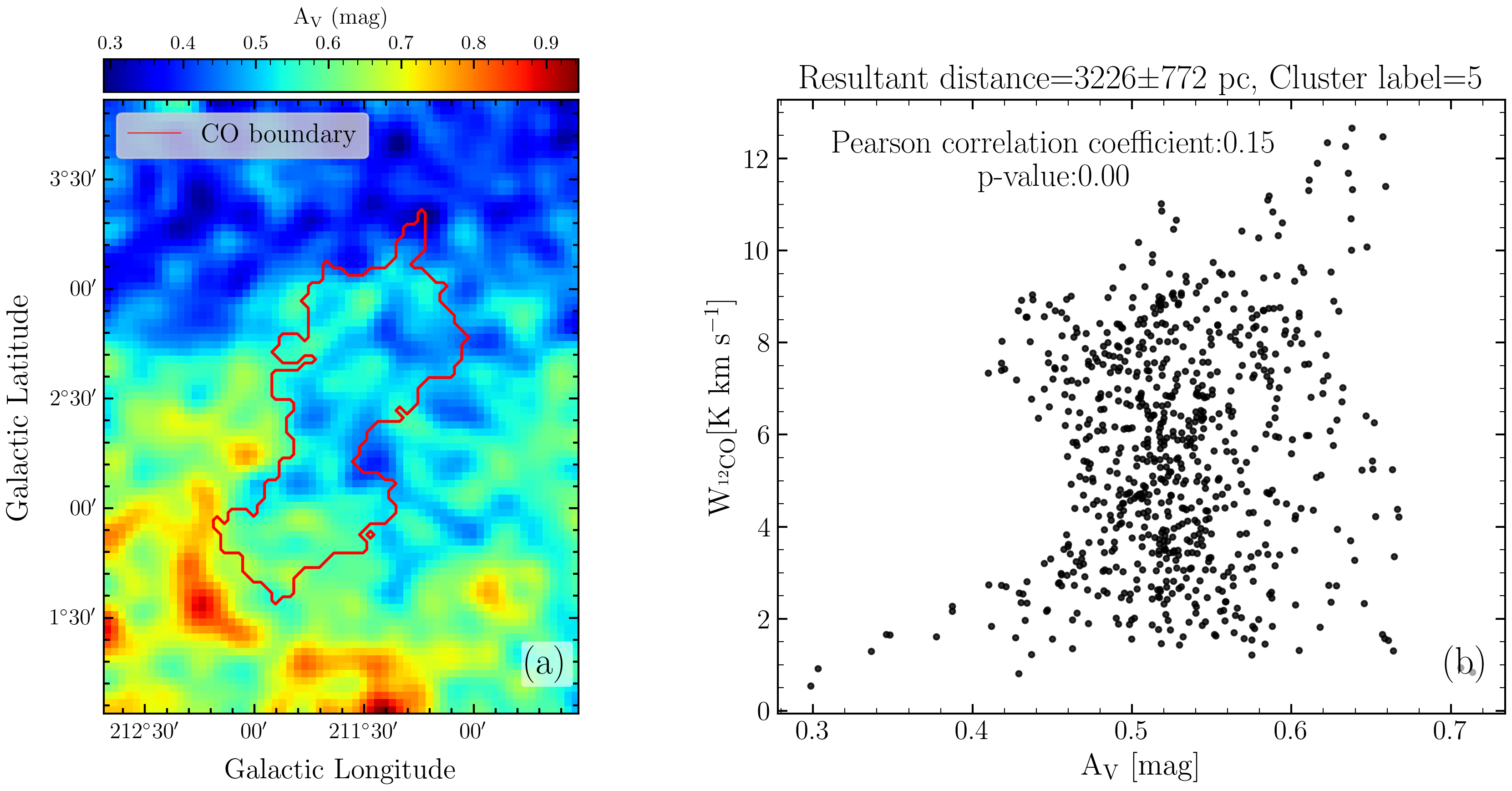}
    \end{minipage}  
    \hfill
    \begin{minipage}{0.49\textwidth}
        \centering
        \includegraphics[width=\textwidth]{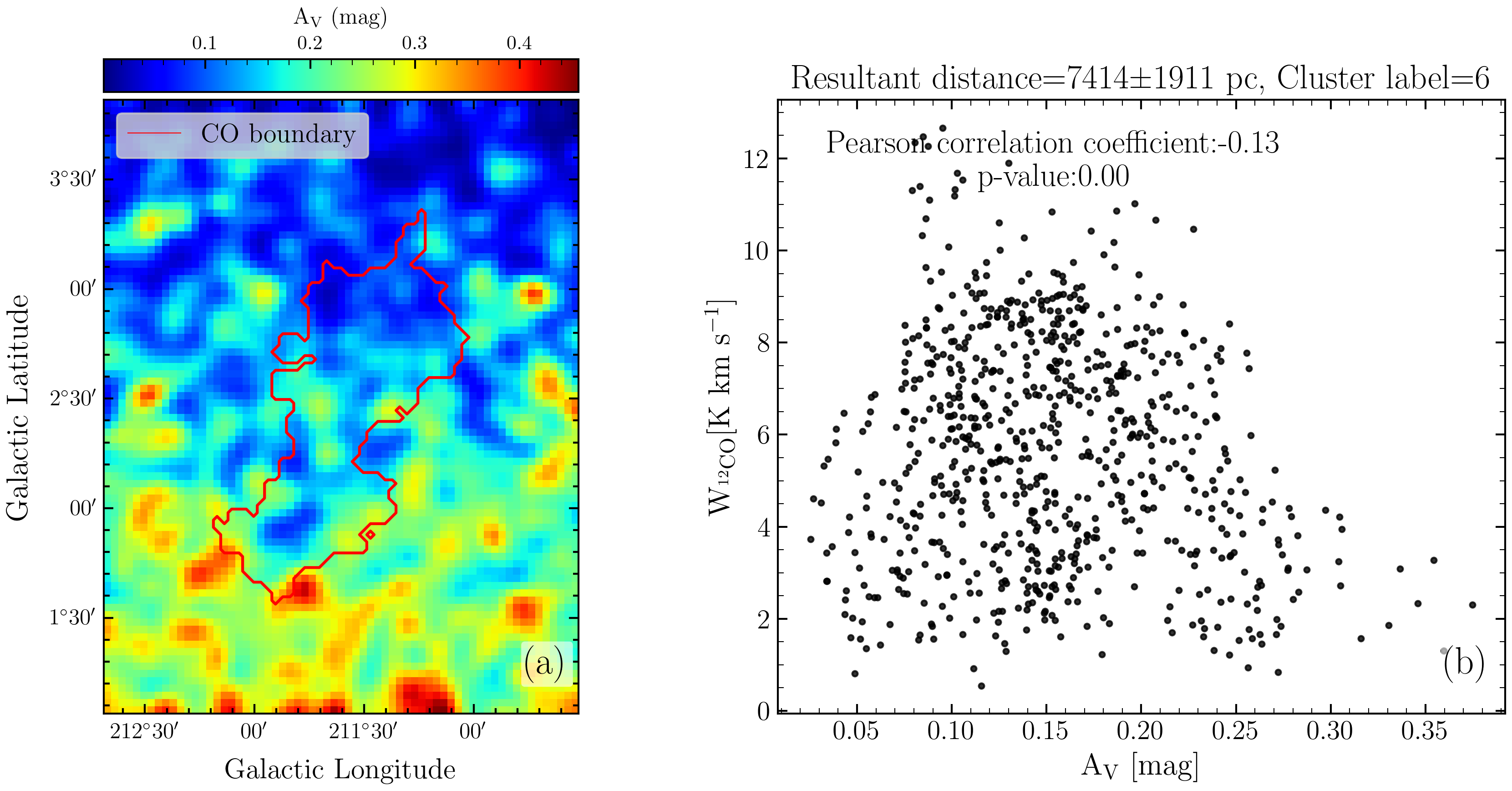}
    \end{minipage}  
    \caption{Reconstructed extinction maps from K-means clustering results and their Pearson correlation with $^{12}$CO integrated intensity maps. Columns 1 and 3 display the reconstructed dust extinction maps. Columns 2 and 4 show the corresponding maps of the Pearson correlation between A$_{\rm V}$ and W$_{\mathrm{CO}}$ at the resultant distance for each cluster derived from the K-means clustering. The result marked with a red frame exhibits morphology most similar to that of the $^{12}$CO integrated intensity map. }
    \label{reconstructed_avmaps}
\end{figure*}

To identify the correct distance among these candidates, we reconstruct 2D extinction maps and compare them with the molecular gas. For each candidate distance, we generate an extinction map by integrating the 3D dust data within a distance interval of $\pm 20\%$ (Figure \ref{reconstructed_avmaps}). We then visually compare the morphology of these reconstructed maps with the $^{12}$CO integrated intensity image. This visual assessment is supported quantitatively by calculating the pixel-by-pixel Pearson correlation coefficient between the extinction ($A_{\mathrm{V}}$) and the CO intensity (W$_{\mathrm{CO}}$). Following the same criteria used in the previous method, we select the candidate that shows the strongest morphological similarity and positive correlation with the CO emission. In Figure \ref{reconstructed_avmaps}, for example, the cluster marked with a red frame exhibits the closest match to the $^{12}$CO map and is adopted as the final cloud distance.

\subsection{Distance-sliced dust map analysis for cloud distances}
\label{sect:method3}

The second method described in Sect.~\ref{sect:method2} relies on detecting distinct steps in the extinction-distance profile to identify dust clouds. However, this approach becomes ineffective when the profile lacks sharp variations. This often occurs with distant clouds that have intrinsically low extinction, or in cases where the number of on-cloud sightlines is too small to reveal statistically significant peaks.

To address these cases, we developed a third approach that determines the distance based solely on the spatial correlation between the dust and the gas. This method involves calculating the pixel-by-pixel Pearson correlation coefficient between the extinction ($A_{\mathrm{V}}$) and the $^{12}$CO integrated intensity (W$_{\mathrm{CO}}$) across a range of distances. We define the molecular cloud distance as the location where this correlation coefficient reaches its maximum value. To ensure reliability, we accept the result only after visually confirming the morphological consistency between the extinction map and the CO emission.

\begin{figure*}[!htbp]
    \centering
    \includegraphics[width=0.85\linewidth]{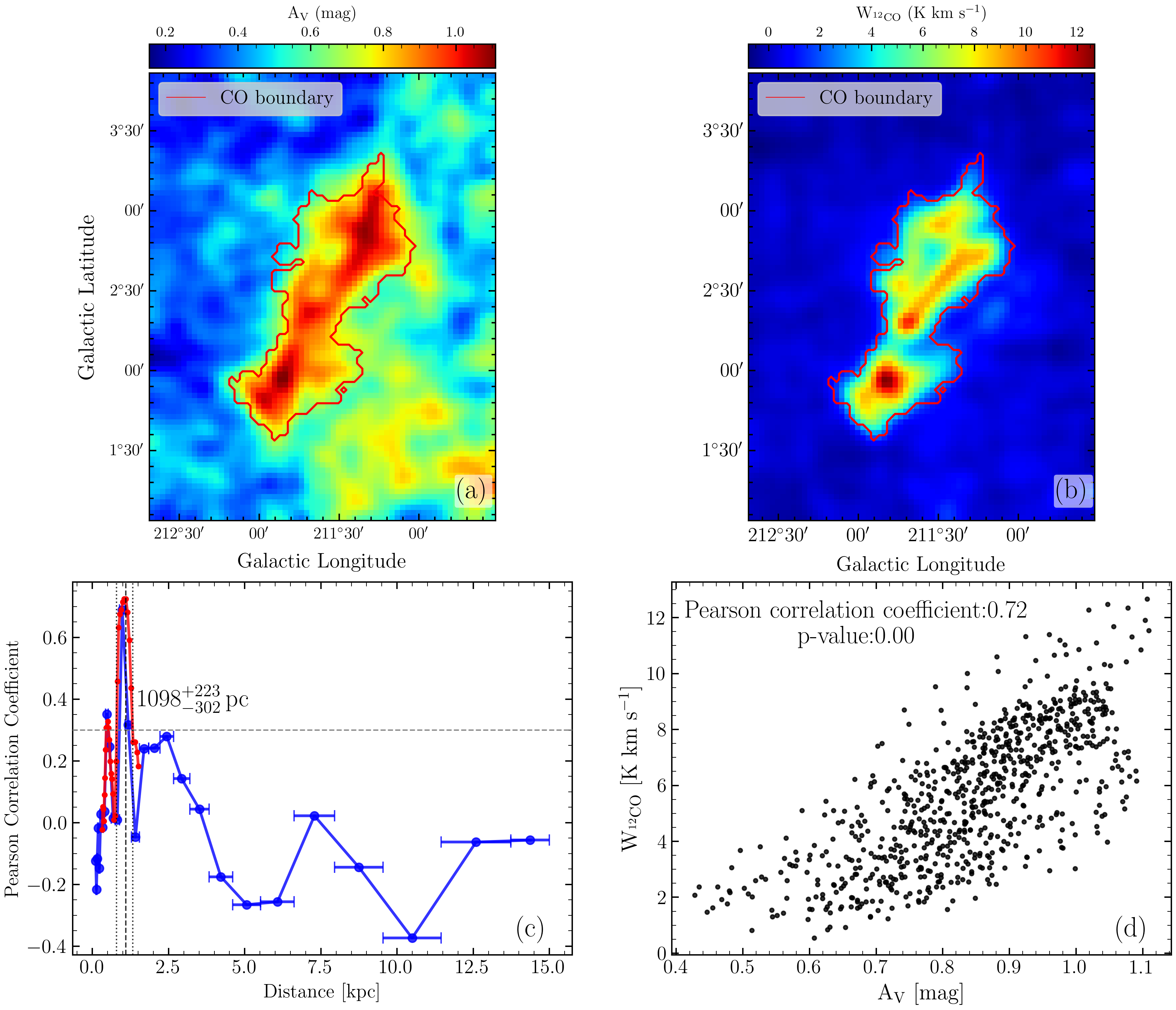}
    \caption{Distance of MWISP G211.613+02.405+007.49 based on the A$_{\rm V}$ vs. W$_{\mathrm{CO}}$ Pearson correlation from distance-sliced dust map. Panels (a), (b), and (d) are identical to those in Figure \ref{distance_kstest} and are described in detail in its caption. (c) the variation of the correlation coefficient at different distance slices. The blue line represents the correlation variation calculated with a 20$\%$ distance interval, while the red line shows a finer 5$\%$ interval. The black solid line indicates the estimated distance and the black dashed lines depict the uncertainty of the cloud distance. 
    }
    \label{distance_correlation}
\end{figure*}

Figure \ref{distance_correlation} illustrates the principle of this correlation-based method. We systematically construct 2D extinction maps by integrating the 3D dust data within a $\pm 20\%$ range around a series of trial distances. 
We then estimate the correlation of 2D extinction maps with CO map for each trial distance. As shown in the correlation profile (Figure \ref{distance_correlation}c), the coefficient peaks at a specific distance, which indicates the physical location of the cloud. This quantitative peak corresponds to the distance where the spatial structures of the dust (Figure \ref{distance_correlation}a) and the molecular gas (Figure \ref{distance_correlation}b) align most closely, resulting in a strong positive correlation (Figure \ref{distance_correlation}d).

\section{Result and discussion}   
\subsection{Distances of the molecular clouds}

Due to the limitations of \textit{Gaia} parallax precision and the resolution of current extinction maps, distance measurements for molecular clouds beyond 3 kpc suffer from significant uncertainties and reduced accuracy. Consequently, we restrict our final catalog to clouds within a heliocentric distance of 3~kpc. Within this volume, we have successfully estimated distances for a total of 1,573 molecular clouds. The results are summarized in Table~\ref{Catalog}. The columns are organized as follows: (1) the catalog index from \cite{YangMWISP+2026}; (2) the molecular cloud name including average $l$-$b$-$V$ coordinates; (3) angular area; (4) linear radii; (5) distance estimated by the statistical method; (6) distance estimated by the morphological matching method; (7) distance estimated by the correlation method; (8) molecular cloud mass, derived assuming a $^{12}$CO-to-H$_2$ conversion factor of $X=2.0 \times 10^{20}$~cm$^{-2}$~(K~km~s$^{-1}$)$^{-1}$ \citep{Bauermeister+2013}; (9) kinematic distance derived from the \cite{Reid+2014} model; (10) kinematic distance derived from the \cite{Reid+2019} model; and (11) notes. As a representative example, we provide the detailed measurement results for the cloud MWISP G211.613+02.405+007.49. The complete catalog and accompanying diagnostic figures for each cloud are publicly available at ScienceDB (DOI: 10.57760/sciencedb.34703).

\begin{figure*}[!htbp]
    \centering
	\begin{minipage}{0.46\linewidth} 
        \centering
        \includegraphics[width=\linewidth]{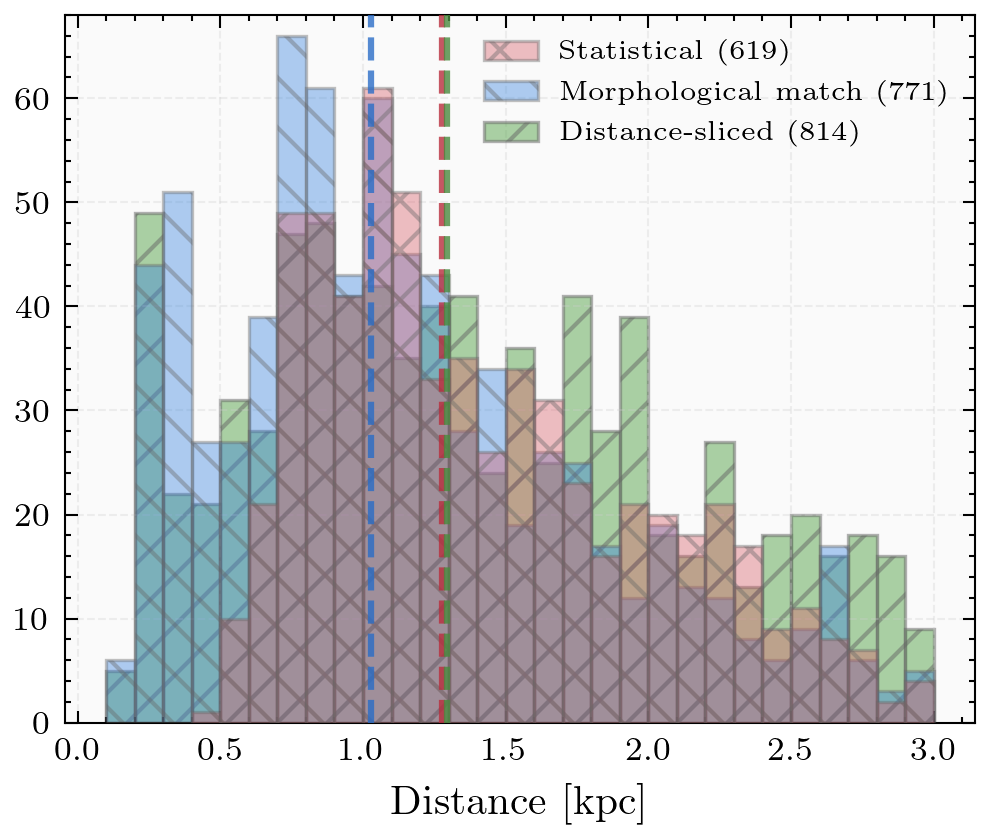}
        \label{Hist_rcpc}
    \end{minipage}
    \hfill 
    \begin{minipage}{0.52\linewidth} 
        \centering
        \includegraphics[width=\linewidth]{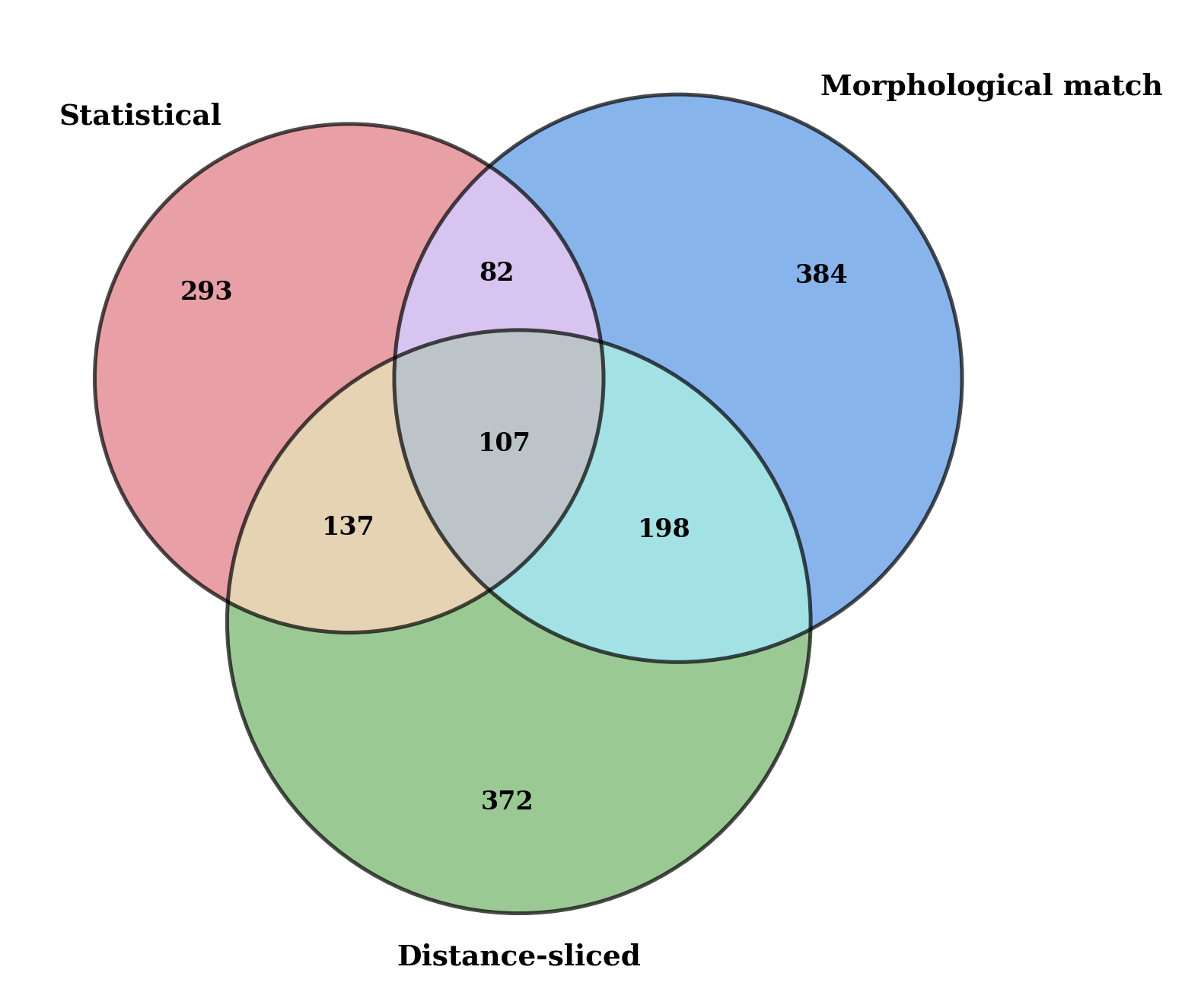}
        \label{Hist_mass}
    \end{minipage}
    \caption{Histogram of distances (left) and Venn diagram (right) for cataloged molecular clouds measured by three independent methods. Colors distinguish the methods: statistical (red), morphological match (blue), distance-sliced (green). Vertical dashed lines in the histogram indicate the median distance for each method, and numbers in the Venn diagram segments show the count of clouds identified by the corresponding combination of methods.
    }
    \label{Venn_methods_hist_distances}
\end{figure*}

The derived distances in the catalog range from 151 pc to 2998 pc, with a typical statistical uncertainty of $\sim 20\%$ and an estimated systematic uncertainty of $\sim 10\%$. 
The performance of the three individual methods is detailed in the left panel of Figure~\ref{Venn_methods_hist_distances}. The statistical method (Method 1) measured 619 clouds (293 exclusively) over a distance range of 494--2996 pc, achieving the highest precision with a typical uncertainty of $\sim 13\%$. The morphological matching method (Method 2) measured 771 clouds (384 exclusively) ranging from 164 to 2971 pc, with a typical uncertainty of $\sim 18\%$. The distance-sliced correlation method (Method 3) measured 814 clouds (372 exclusively) spanning 151-2998 pc, with a typical uncertainty of $\sim 20\%$. As visualized in the Venn diagram (Figure~\ref{Venn_methods_hist_distances}, right), there is substantial overlap between the results: 107 clouds were measured by all three methods. 
The distances of 417 molecular clouds were measured by two of the three methods, with results agreeing within a median discrepancy of 10\%. Among these, 82 clouds were measured by the combination of Methods 1 and 2, 137 by Methods 1 and 3, and 198 by Methods 2 and 3. This concordance provides a robust basis for cross-validation, while the discrepancies highlight the specific regimes of applicability for each technique.

\begingroup
\renewcommand{\arraystretch}{1.3}         
\setlength{\tabcolsep}{2.0 pt}
\begin{longtable}{TTTTTTTTTTTT}
    \caption{\normalsize Distances to 1573 molecular clouds.}\label{Catalog} \\
    \hline
    \hline
    cloudIndex$^{\textcolor{blue}{1}}$ & cloudName & Area & r$^{\textcolor{blue}{2}}$ & D$_{\rm KStest}$$^{\textcolor{blue}{3}}$ & D$_{\rm Kmeans}$$^{\textcolor{blue}{4}}$ & D$_{\rm Corrlation}$$^{\textcolor{blue}{5}}$ & Mass$^{\textcolor{blue}{6}}$ & D$_{\rm Reid2014}$$^{\textcolor{blue}{7}}$ & D$_{\rm Reid2019}$$^{\textcolor{blue}{8}}$ & Notes  \\
    & & (deg$^2$) & (pc) & (pc) & (pc) & (pc) & (10$^{3}$ M$_{\odot}$) & (kpc) & (kpc) &  \\
    (1) & (2) & (3) & (4) & (5) & (6) & (7) & (8) & (9) &(10) &(11)\\
    \hline
    2868	&	MWISP G$014.138+02.518+015.57$	&	0.60	&	9.82	&							&	1591	$^{+	321	}_{-	321	}$	&	1279	$^{+	82	}_{-	58	}$	&	5.15	&	1.8	$^{+	1.23	}_{-	0.93	}$	&	1.15	$^{+	0.14	}_{-	0.14	}$	&	L347	\\
    4908	&	MWISP G$016.897-02.428+017.91$	&	0.44	&	16.68	&							&							&	2547	$^{+	438	}_{-	700	}$	&	74.61	&	1.81	$^{+	1.04	}_{-	0.82	}$	&	1.14	$^{+	0.13	}_{-	0.13	}$	&	L379	\\
    5560	&	MWISP G$017.714+03.072+010.30$	&	0.10	&	3.67	&	1182	$^{+	118	}_{-	118	}$	&							&							&	1.29	&	1.02	$^{+	0.51	}_{-	0.95	}$	&	1.19	$^{+	0.13	}_{-	0.13	}$	&	L386	\\
    9442	&	MWISP G$023.309-01.131+038.92$	&	0.06	&	5.24	&							&							&	2271	$^{+	346	}_{-	530	}$	&	2.74	&	2.92	$^{+	0.62	}_{-	0.55	}$	&	3.54	$^{+	0.38	}_{-	0.38	}$	&	L455	\\
    12591	&	MWISP G$027.893-02.163+018.05$	&	1.84	&	28.40	&							&	2184	$^{+	381	}_{-	381	}$	&	2074	$^{+	230	}_{-	519	}$	&	89.07	&	1.38	$^{+	0.79	}_{-	0.69	}$	&	1.31	$^{+	0.75	}_{-	0.75	}$	&	L523	\\
    13386	&	MWISP G$029.058+02.262+032.83$	&	0.77	&	14.14	&							&	1472	$^{+	249	}_{-	249	}$	&	1729	$^{+	134	}_{-	254	}$	&	20.54	&	2.35	$^{+	0.65	}_{-	0.59	}$	&	4.31	$^{+	2.15	}_{-	2.15	}$	&	L541	\\
    23910	&	MWISP G$042.786-02.817+035.36$	&	0.25	&	6.39	&							&	1248	$^{+	212	}_{-	212	}$	&	1472	$^{+	464	}_{-	448	}$	&	2.31	&	2.5	$^{+	0.69	}_{-	0.71	}$	&	1.97	$^{+	0.57	}_{-	0.57	}$	&	L647	\\
    25121	&	MWISP G$044.758+04.033+020.01$	&	0.82	&	9.90	&	1150	$^{+	35	}_{-	35	}$	&	778	$^{+	133	}_{-	133	}$	&	634	$^{+	298	}_{-	114	}$	&	4.17	&	1.46	$^{+	0.73	}_{-	0.72	}$	&	0.58	$^{+	0.2	}_{-	0.2	}$	&	L662	\\
    26207	&	MWISP G$046.873-03.034+026.76$	&	0.47	&	9.30	&	1308	$^{+	108	}_{-	108	}$	&	1785	$^{+	346	}_{-	346	}$	&	1472	$^{+	183	}_{-	159	}$	&	4.40	&	1.99	$^{+	0.75	}_{-	0.78	}$	&	1.53	$^{+	0.55	}_{-	0.55	}$	&	L682	\\
    26903	&	MWISP G$048.241-01.705+021.59$	&	3.04	&	20.57	&							&	1354	$^{+	254	}_{-	254	}$	&	1090	$^{+	170	}_{-	252	}$	&	25.00	&	1.64	$^{+	0.76	}_{-	0.79	}$	&	1.21	$^{+	0.55	}_{-	0.55	}$	&	L698	\\
    28660	&	MWISP G$051.315+03.322+015.49$	&	3.04	&	12.25	&	684	$^{+	56	}_{-	56	}$	&	961	$^{+	164	}_{-	164	}$	&	776	$^{+	370	}_{-	300	}$	&	13.84	&	1.24	$^{+	0.8	}_{-	0.84	}$	&	0.83	$^{+	0.56	}_{-	0.56	}$	&	L714	\\
    29236	&	MWISP G$052.181+01.580+023.80$	&	3.78	&	17.93	&							&	848	$^{+	122	}_{-	122	}$	&	1239	$^{+	101	}_{-	353	}$	&	24.02	&	1.97	$^{+	0.85	}_{-	0.97	}$	&	1.42	$^{+	0.61	}_{-	0.61	}$	&	L716	\\
    34277	&	MWISP G$062.121-04.484+014.14$	&	0.13	&	4.73	&							&							&	1348	$^{+	549	}_{-	333	}$	&	1.49	&	1.56	$^{+	1.11	}_{-	1.11	}$	&	0.95	$^{+	0.98	}_{-	0.98	}$	&	L797	\\
    39214	&	MWISP G$070.689-03.953+005.75$	&	0.3	&	8.05	&							&	1431	$^{+	234	}_{-	234	}$	&	1686	$^{+	343	}_{-	426	}$	&	2.64	&	0.99	$^{+	0.99	}_{-	4.8	}$	&	1.59	$^{+	0.04	}_{-	0.04	}$	&	L838	\\
    40441	&	MWISP G$072.576+01.304+002.54$	&	0.12	&	6.97	&							&							&	2027	$^{+	563	}_{-	188	}$	&	4.04	&	0.56	$^{+	0.56	}_{-	5.09	}$	&	3.75	$^{+	0.89	}_{-	0.89	}$	&	L856	\\
    44328	&	MWISP G$078.715+04.014-002.60$	&	0.15	&	5.79	&							&							&	1514	$^{+	650	}_{-	117	}$	&	7.02	&	3.33	$^{+	1.43	}_{-	1.43	}$	&	1.51	$^{+	1.24	}_{-	1.24	}$	&	L891	\\
    ...	&	...	&	...	&	...	& ... &	...	&	...	&	...	&	...	&	...	&	...	\\
    
    \hline\noalign{\smallskip}
    \multicolumn{12}{l}{
    \footnotesize \textbf{Notes.} 
    \parbox[t]{0.9\textwidth}{\footnotesize
    $^1$ The index in the catalog of \cite{YangMWISP+2026}.
    $^2$ The linear radii $r$ are calculated by the distance, which is the uncertainty-weighted mean from the our methods.
    $^3$ The estimated distances by the statistical method based on extinction–distance distribution differences. 
    $^4$ The estimated distances by the morphological matching method. 
    $^5$ The estimated distances by the A$_{\rm V}$ vs. $^{12}$CO correlation of distance-sliced dust map. 
    $^6$ Total mass of molecular gas in molecular clouds estimated with the $^{12}$CO-to-H2 mass conversion factor of X=2.0$\times$10$^{20}$ cm$^\mathrm{-2}$ (K km s$^{-1}$)$^{-1}$ \citep{Bauermeister+2013}, which only takes CO-bright components into account. 
    $^7$ Kinematic distances derived from the A5 model of \citet{Reid+2014}. 
    $^8$ Kinematic distances derived from the model in \citet{Reid+2019}. 
    For cloud MWISP G014.938-00.120+123.30, the distance estimated in this work may be unreliable due to its exceptionally high $V_{\rm LSR}$, and should therefore be used with caution. For comparison, the kinematic distance derived from \citet{Reid+2019} is as large as 8 kpc, much larger than our estimate.
    }
    } \\
\end{longtable}
\endgroup

These variations in sample size, coverage, and uncertainty stem directly from the inherent strengths and limitations of each approach. The statistical method yields the smallest uncertainties because it relies on precise star-by-star data from the StarHorse catalog \citep{Anders+2022} to detect extinction jumps. However, this requirement restricts its use to clouds with sufficient foreground and background stars, making it unreliable for very nearby or distant sources. 
In contrast, the morphological matching and distance-sliced methods both utilize the 3D dust map of \citet{Green+2019}, allowing them to probe a larger number of clouds, particularly at closer distances, albeit with slightly higher uncertainties inherent to dust decomposition. The distance uncertainties in the \citet{Green+2019} catalog are dominated by Gaia parallax measurements (typically 3\%–15\%), comparable to the statistical uncertainties of our methods (13\%–20\%). The median discrepancy of about 10\% from cross-validation among our three methods suggests that systematic misassignments affect only a small fraction of clouds, primarily in complex regions such as the inner Galaxy. Specifically, the morphological matching method depends on detecting distinct peaks in line-of-sight extinction profiles; it therefore becomes ineffective for small clouds with too few sightlines to establish a reliable match. The distance-sliced method circumvents this by determining the distance where the spatial correlation between extinction and CO intensity is maximized. This allows it to successfully measure smaller clouds that lack sufficient sightlines for profile fitting, although it remains susceptible to failure in regions with severe foreground dust contamination.

\begin{table}[htbp]
\centering
\renewcommand{\arraystretch}{1.3}
\setlength{\tabcolsep}{2.1mm}{
\caption{Cloud counts and total flux at each selection step}
\label{table_cloud_selection_num_flux}
\begin{tabular}{ccccl}
\hline
Step & Description & Number of Clouds & Total flux (K~km~s$^{-1}$~arcmin$^2$) \\
(1) & (2) & (3) & (4) \\
\hline
1 & MWISP $^{12}$CO survey & - & $7.69 \times 10^{7}$ \\
2 & Complete MWISP cloud catalog & 103,517 & $1.18 \times 10^{7}$ (15.3\%) \\
3 & After angular area threshold ($\geq 0.01$ deg$^{2}$) & 10,929 & $1.06 \times 10^{7}$ (13.7\%) \\
4 & After distance measurement & 1,573 & $5.96 \times 10^{6}$ (7.8\%) \\
\hline
\multicolumn{4}{l}{
\footnotesize \textbf{Notes.} 
\parbox[t]{0.9\textwidth}{\footnotesize
Percentages in parentheses indicate the fraction of the total MWISP $^{12}$CO flux (relative to Step 1) retained at each step.
}
} \\
\end{tabular}}
\end{table}

We examine the number of clouds and total flux retained at each step of the selection process, as summarized in Table~\ref{table_cloud_selection_num_flux}. The total $^{12}$CO flux of the MWISP survey is $7.69 \times 10^7$ K km s$^{-1}$ arcmin$^{2}$, of which 82\% is recovered by the DBSCAN algorithm \citep{YangMWISP+2026}. In the construction of the cloud catalog, incomplete clouds near survey boundaries were removed. The resulting complete MWISP cloud catalog contains 103,517 clouds with a total flux of $1.18 \times 10^7$ K km s$^{-1}$ arcmin$^2$, retaining 15.3\% of the total MWISP flux. After applying the angular area threshold of 0.01 deg$^2$ (Section~2.1), 10,929 clouds remain, retaining 13.7\% of the total MWISP flux. Ultimately, distances are derived for 1,573 clouds, which retain 7.8\% of the total MWISP flux, corresponding to approximately 51\% of the flux in the complete MWISP cloud catalog.

\begin{figure*}[!htbp]
    \centering
	\includegraphics[width=0.65\linewidth]{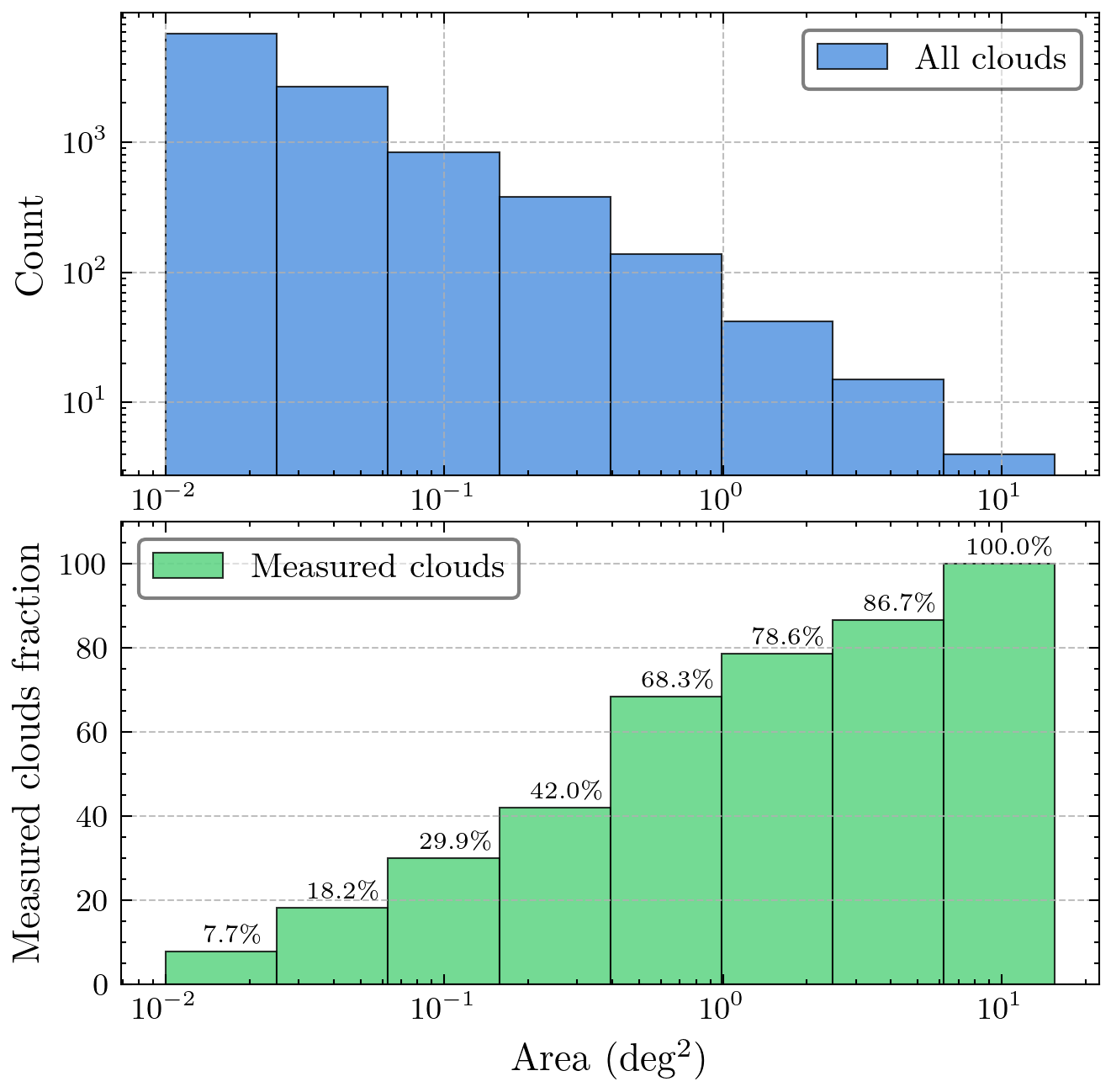}
    \caption{Angular area distribution of molecular clouds, with the top panel showing the number of all clouds and the bottom panel showing the fraction of clouds with distance measurements.
    }
    \label{Hist_angular_area}
\end{figure*}

To assess the completeness of our catalog, we then compare the angular area distribution of the distance-measured subsample with that of the full molecular cloud population (Figure~\ref{Hist_angular_area}). The top panel of Figure~\ref{Hist_angular_area} shows the number of all clouds in each angular area bin for the full sample, while the bottom panel of Figure~\ref{Hist_angular_area} presents the fraction of clouds with successful distance measurements in each bin. The results show that distances have been successfully measured for more than 80\% of MCs with angular areas greater than 1 deg$^2$ using our methods. In addition, about 20\% of MCs with angular areas smaller than 0.1 deg$^2$ are also measured, due to our combined strategies. Among these smaller MCs, about 90\% of them are measured for the first time.

\subsubsection{Comparison of the three distance estimation methods}

\begin{figure*}[!htbp]
    \centering
	\includegraphics[width=1\linewidth]{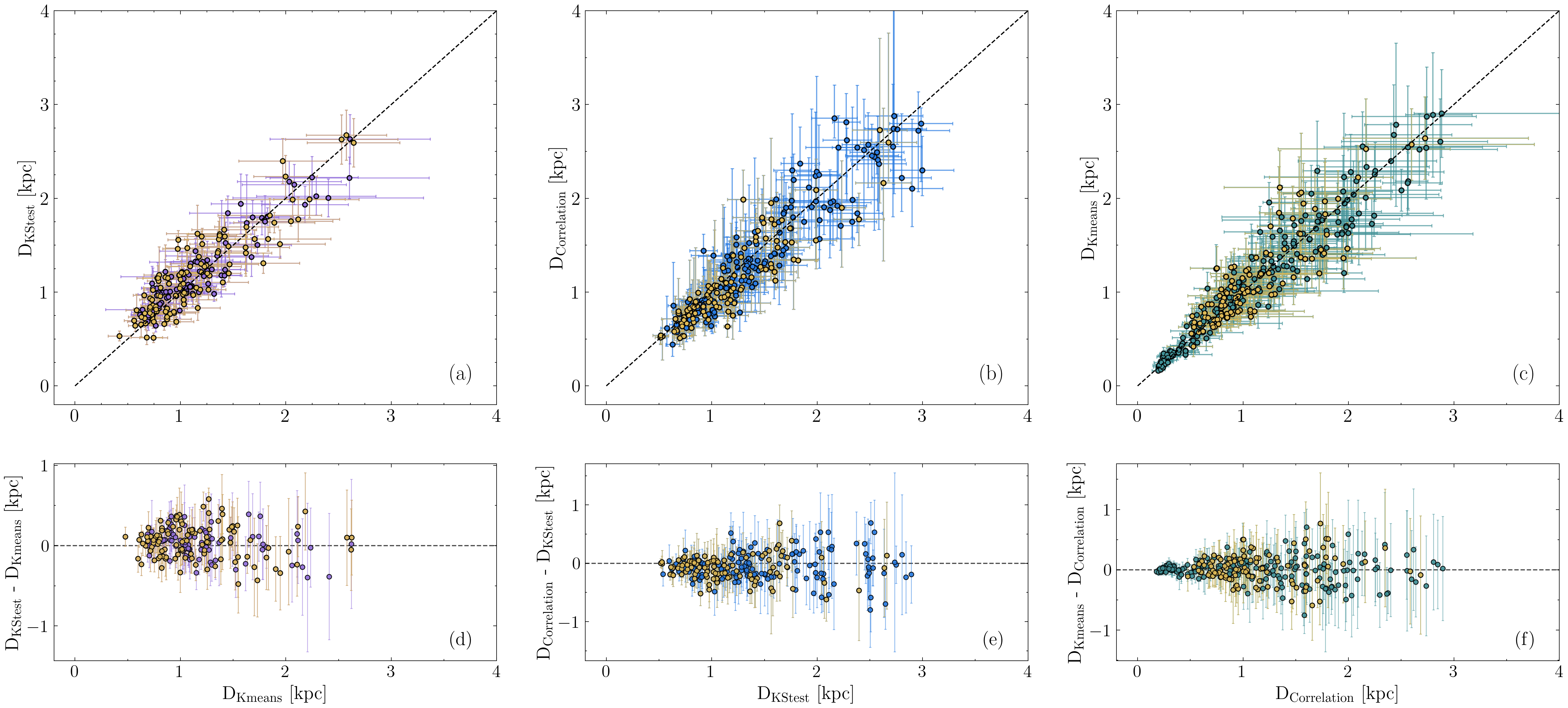}
    \caption{Comparison of molecular cloud distances derived using three independent methods in this work: statistical method based on extinction–distance distribution differences (D$_{\rm {KStest}}$), morphological matching method (D$_{\rm {Kmeans}}$), and correlation analysis of distance‑sliced dust maps (D$_{\rm {Correlation}}$). The top and bottom rows show pairwise comparisons and their corresponding difference plots, respectively: (a, d) D$_{\rm {KStest}}$ vs. D$_{\rm {Kmeans}}$; (b, e) D$_{\rm {Correlation}}$ vs. D$_{\rm {KStest}}$; (c, f) D$_{\rm {Correlation}}$ vs. D$_{\rm Kmeans}$. In all panels, molecular clouds with distances available from three methods are highlighted in yellow. The dashed lines indicate the one‑to‑one relation (top row) and the zero‑difference line (bottom row).}
    \label{compare_methods}
\end{figure*}

To evaluate the robustness of our results, we cross-compare the distances derived from the three independent techniques: the statistical KS test ($D_{\rm {KStest}}$), the K-means clustering of dust components ($D_{\rm {Kmeans}}$), and the spatial correlation analysis ($D_{\rm {Correlation}}$). As shown in Figure~\ref{compare_methods}, the measurements obtained from these distinct approaches exhibit strong agreement within their uncertainties.

We quantify this consistency by calculating the median offsets for each pairwise comparison. The resulting differences are 0.061 kpc between $D_{\rm {KStest}}$ and $D_{\rm {Kmeans}}$, $-0.069$ kpc between $D_{\rm {KStest}}$ and $D_{\rm {Correlation}}$, and 0.007 kpc between $D_{\rm {Kmeans}}$ and $D_{\rm {Correlation}}$. At a representative distance of $\sim 1$ kpc, these offsets correspond to relative differences of less than 10\%, which are well within the typical measurement uncertainty of $\sim 20\%$. This demonstrates that there are no significant systematic biases between the methods. Based on this consistency, for clouds where reliable estimates are available from all three methods, we adopt the uncertainty-weighted mean as the final distance to maximize precision.

\begin{figure*}[!htbp]
    \centering
	\includegraphics[width=0.55\linewidth]{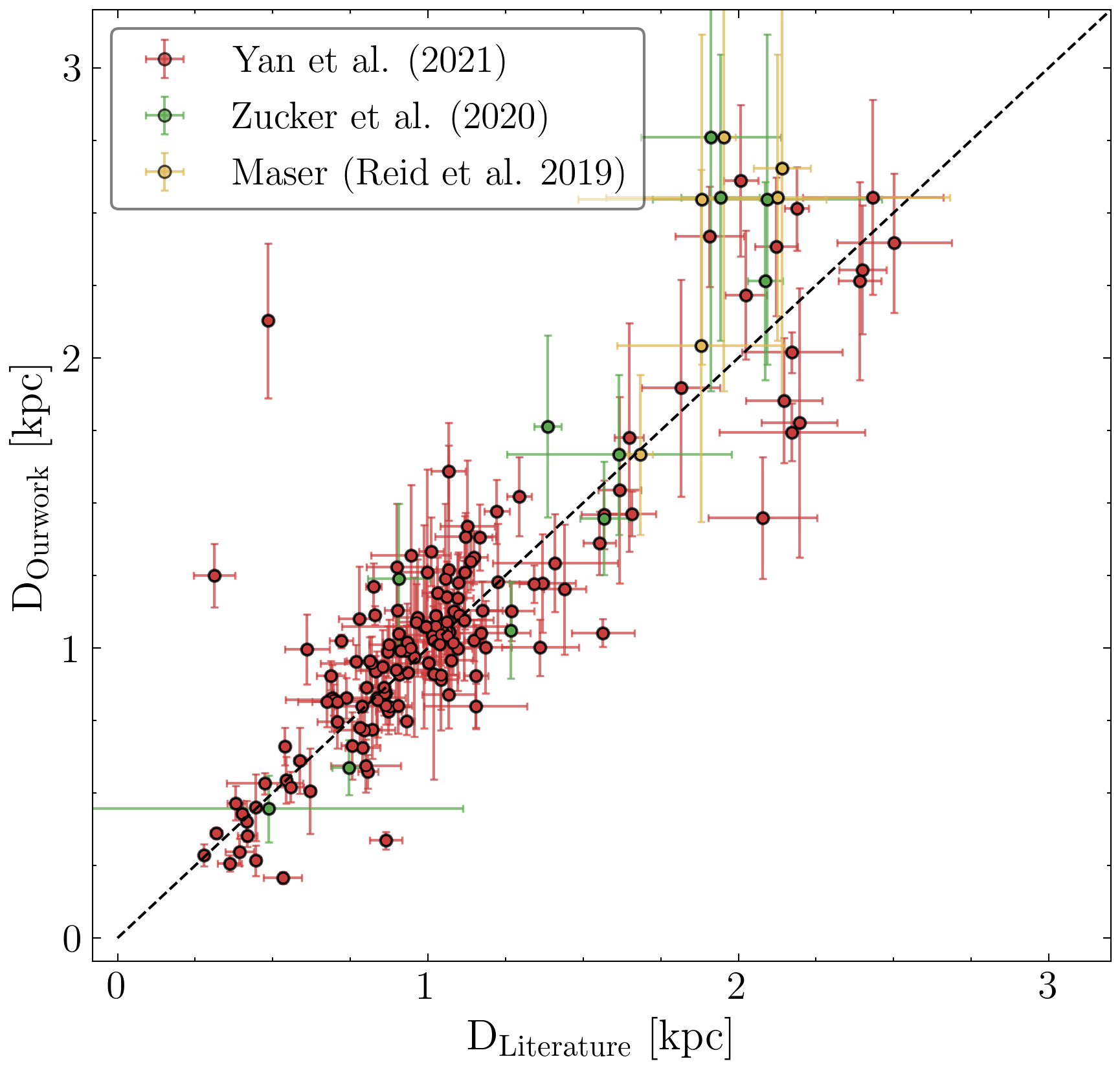}
    \caption{Comparison of molecular cloud distances from this work with previous results.}
    \label{compare_literture}
\end{figure*}

\subsubsection{Comparison with previous results}

To validate the reliability of our measurements, we compare our distance estimates with those reported in previous studies (see Figure~\ref{compare_literture}). Our comparison sample includes 149 molecular clouds from \citet{Yan+2021} (red points), 11 clouds from \citet{Zucker+2020} (green points), and 6 precise maser parallax measurements from \citet{Reid+2019} (yellow points).

Overall, our results show excellent agreement with the literature. The distances from \citet{Zucker+2020} and the maser measurements from \citet{Reid+2019} are consistent with our values within 20\%. Similarly, for the larger sample from \citet{Yan+2021}, the agreement is within 10\% for the vast majority of sources. We identify only two significant outliers in the comparison with \citet{Yan+2021}. A closer inspection of the extinction maps suggests that these discrepancies likely arise from confusion with foreground dust structures, which may have led to underestimated distances in the previous work. Aside from these specific cases, the broad consistency across different datasets confirms that our approach yields robust and reliable distance estimates.

\subsubsection{Comparison with kinematic distances}

\begin{figure*}[!htbp]
    \centering
    \includegraphics[width=1\linewidth]{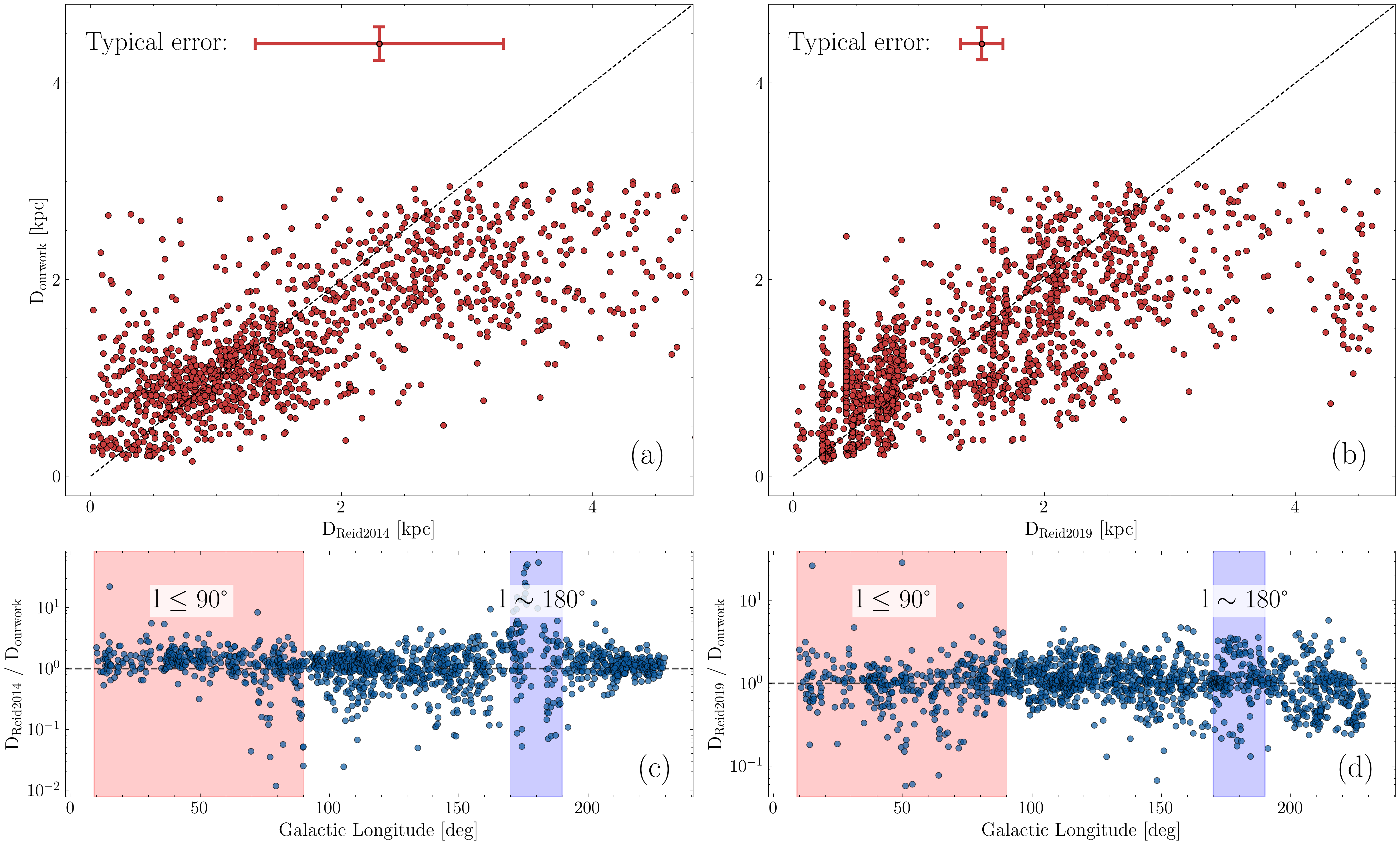}
    \caption{Comparison with kinematic distances derived from the A5 model in \cite{Reid+2014} and its improved version from \cite{Reid+2019}. In panels (a) and (b), typical errors for the two datasets are indicated in the upper-left corners: 0.171~kpc (this work) and 0.99~kpc \citep{Reid+2014} in panel (a), and 0.164~kpc (this work) and 0.170~kpc \citep{Reid+2019} in panel (b). Panels (c) and (d) plot \(D_{\mathrm{Reid}}/D_{\mathrm{thiswork}}\) versus Galactic longitude for the \cite{Reid+2014} and \cite{Reid+2019} datasets, respectively. The red and blue shaded regions indicate the longitude ranges \(l \leq 90^{\circ}\) and \(l \sim 180^{\circ}\), respectively. The horizontal dashed line marks \(D_{\mathrm{Reid}}/D_{\mathrm{thiswork}} = 1\).}
    \label{compare_KD}
\end{figure*}

Figure~\ref{compare_KD} compares our distance estimates with kinematic distances derived using the Galactic rotation curve model (A5) from \citet{Reid+2014} and its updated version from \citet{Reid+2019}, as shown in panels (a) and (b). For this comparison, we adopt the ``near'' kinematic distance solution from the \citet{Reid+2019} model. While a general linear correlation is evident, the data exhibit significant dispersion, which is a known characteristic of kinematic distances due to the influence of peculiar motions.

Notably, systematic offsets are observed for clouds located near Galactic longitudes of $l \lesssim 90\degr$ and $l \sim 180\degr$, as illustrated in panels (c) and (d) of Figure Figure~\ref{compare_KD}. These deviations are attributable to the intrinsic geometric and dynamic limitations of the kinematic method in these specific directions. As discussed in \citet{Reid+2019}, kinematic distances are unreliable in the inner Galaxy ($l \lesssim 90\degr$) due to the near-far ambiguity and velocity crowding near tangent points, particularly toward the Galactic center and in the directions of l $\sim$ 70–90 degrees. Similarly, toward the Galactic anticenter ($l \sim 180\degr$), the line-of-sight velocity projection approaches zero, causing the velocity-distance relation to become degenerate and rendering kinematic estimates highly uncertain.

In contrast, a fundamental advantage of our approach is its independence from Galactic rotation models. Both our stellar sample and the underlying 3D extinction maps are grounded in \textit{Gaia} parallax measurements, which provide direct geometric constraints. Consequently, our method yields reliable distances even in the inner Galaxy and toward the anticenter, regions where kinematic methods are structurally limited by the geometry of Galactic rotation.

\begin{figure*}[!htbp]
    \centering
	\includegraphics[width=1.0\linewidth]{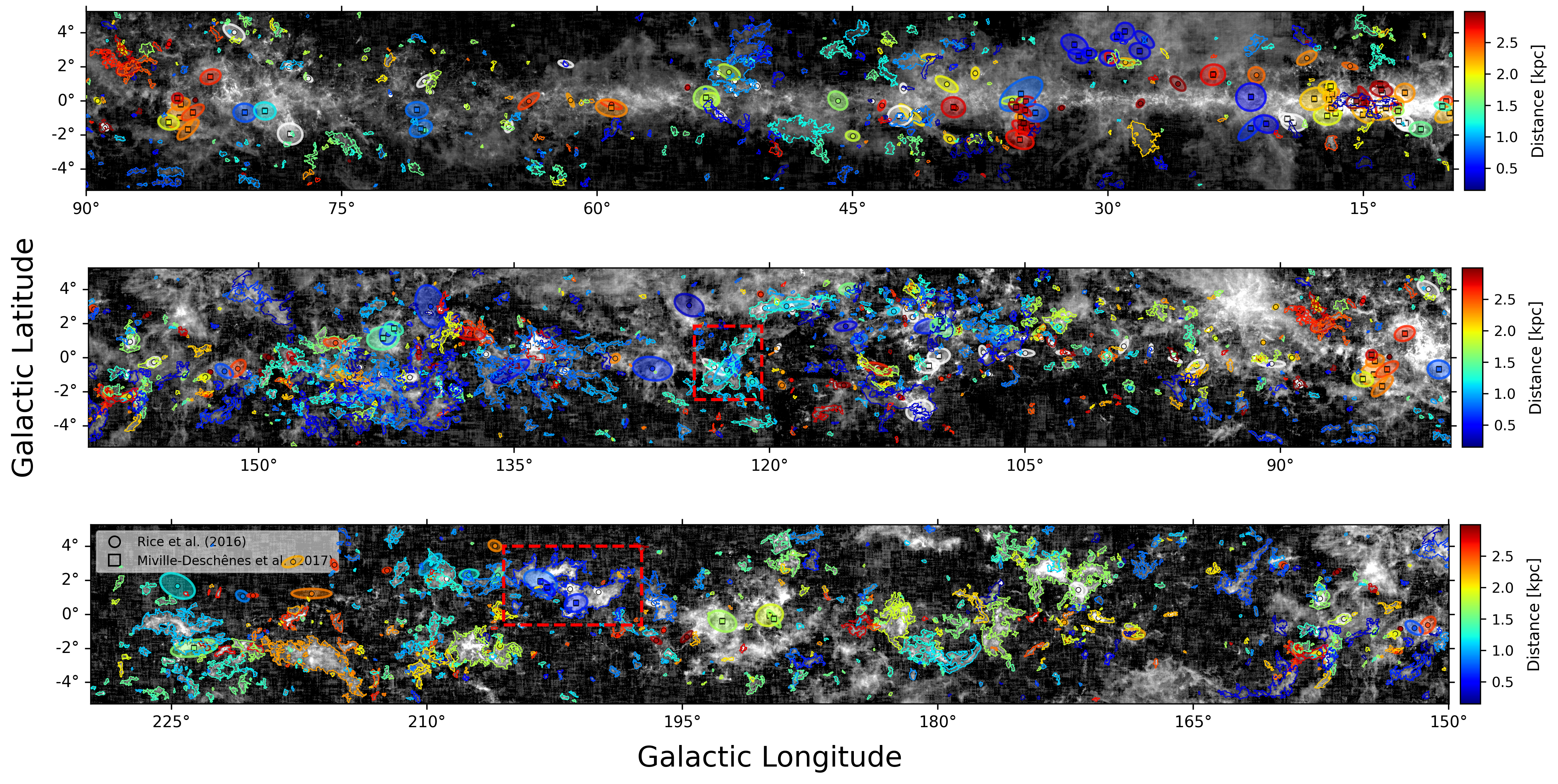}
    \caption{The $l$--$b$ distribution of molecular clouds cataloged in this work. The background is the integrated intensity map of $^{12}$CO emission. Colored contours outline the boundaries of the clouds identified in this study, where the color scale indicates the distance to each cloud. The top, middle, and bottom panels show clouds in the Galactic longitude ranges of $9.75^{\circ}$--$90^{\circ}$, $80^{\circ}$--$160^{\circ}$, and $150^{\circ}$--$229.75^{\circ}$, respectively. For comparison, previously cataloged clouds from \cite{Rice+2016} and \cite{Miville-Deschenes+2017} (both based on kinematic distances) are overplotted as circles and squares, respectively, with ellipses representing their angular sizes and position angles. For these comparison catalogs, clouds with kinematic distances $\le 3$ kpc are shown in color following the same distance scale as the cloud contours, while those with kinematic distances $> 3$ kpc are shown in white.}
    \label{lb_distribution}
\end{figure*}

\subsection{The spatial distribution and physical properties of molecular clouds}

Figure~\ref{lb_distribution} presents the $l$--$b$ distribution of cataloged clouds in this work that have reliable distance measurements. The background is the integrated intensity map of $^{12}$CO emission from MWISP. Colored contours outline the boundaries of our cataloged clouds, with the color scale indicating the distance to each cloud. The three panels cover Galactic longitude ranges of $9.75^{\circ}$--$90^{\circ}$ (top), $80^{\circ}$--$160^{\circ}$ (middle), and $150^{\circ}$--$229.75^{\circ}$ (bottom). For comparison, previously cataloged clouds from \cite{Rice+2016} (218 clouds total: 118 within 3 kpc, 100 beyond 3 kpc) and \cite{Miville-Deschenes+2017} (99 clouds total: 74 within 3 kpc, 25 beyond 3 kpc), both based on kinematic distances, are overplotted as circles and squares, respectively, with ellipses indicating their angular sizes and position angles derived from a brightness-weighted inertia analysis. For these comparison catalogs, clouds with kinematic distances within 3 kpc are shown in color using the same distance scale as the cloud contours, while those with kinematic distances beyond 3 kpc that overlap the cloud contours are shown in white. A spatial cross-matching between our catalog and the two comparison catalogs shows that approximately 61.9\% of the \cite{Rice+2016} clouds and 24.2\% of the \cite{Miville-Deschenes+2017} clouds have counterparts in our sample. As shown in Figure~\ref{lb_distribution}, the correspondence is good in some regions (e.g., $l\sim122^\circ$), but a one-to-many relationship is observed around $l\sim200^\circ$. This discrepancy arises from projection effects in Position-Position-Velocity (PPV) space, where a single Position-Position-Position (PPP) structure with a strong internal velocity gradient can fragment into multiple PPV structures. However, we also find that many clouds from the comparison catalogs do not have counterparts in our sample. This can be attributed to several factors. First, the complete MWISP cloud catalog excludes both large, complex clouds that DBSCAN cannot properly resolve, especially in the inner Galaxy, and clouds located near the edges of the survey that are incomplete. Second, previous catalogs rely on kinematic distances, which suffer from the near/far ambiguity and systematic errors from non-circular motions (see Section~4.1.3), whereas we determine distances by combining MWISP CO maps with 3D extinction maps for higher accuracy.

\begin{figure*}[!htbp]
    \centering
    \includegraphics[width=0.95\linewidth]{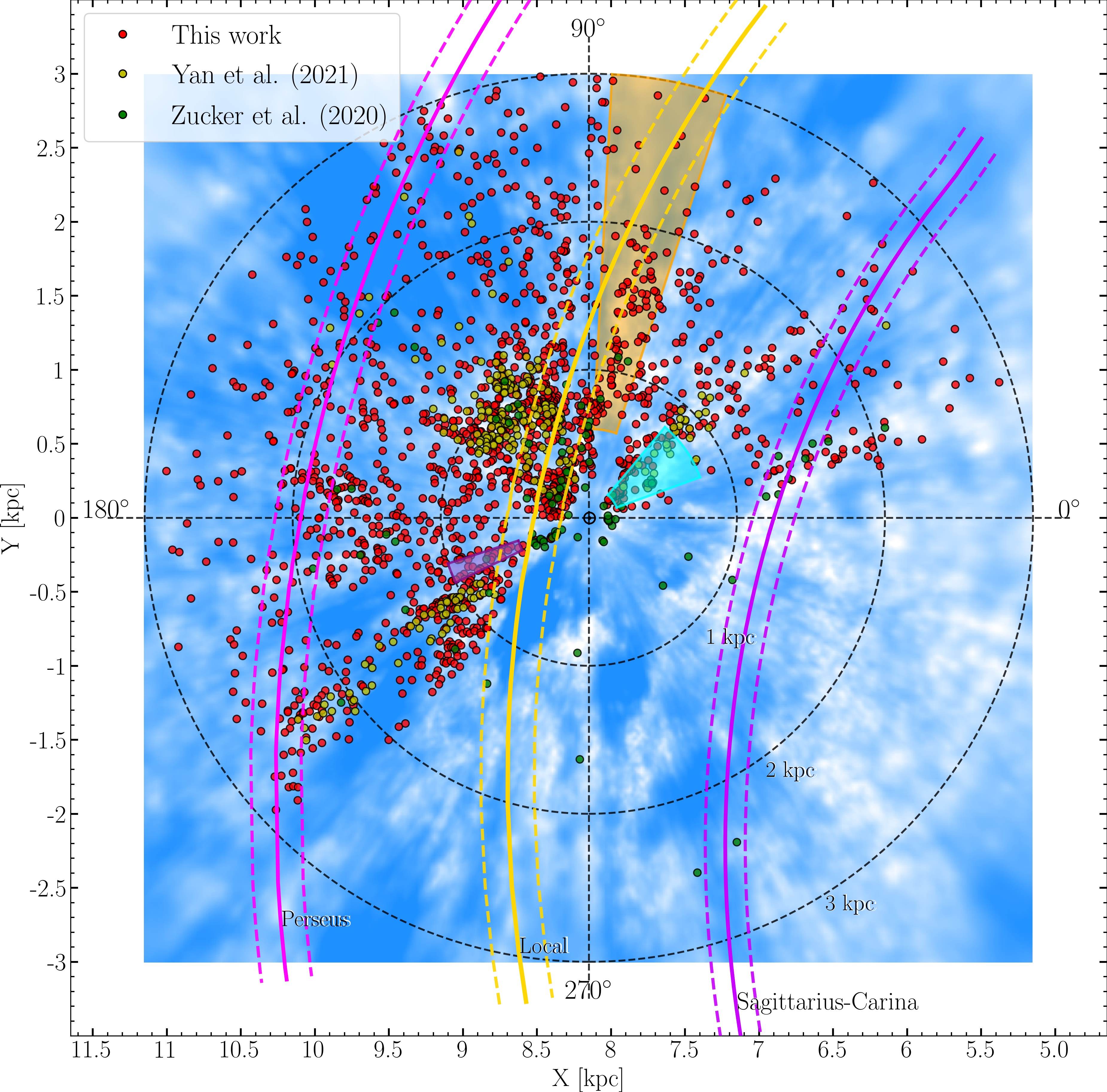}
    \caption{The spatial distribution of the molecular clouds identified in the current work (red hollow circles), \citet[yellow]{Yan+2021} and \citet[green]{Zucker+2020} in the Galactic coordinates. The background is an extinction map from \cite{Vergely+2022} of the Galactic dust of disc vertical height $|Z| < 0.4 \, \text{kpc}$. The Cygnus \citep{Zhang+2024,Zhang+2025}, Aquila Rift \citep{Su+2020}, and Monoceros OB1 \citep{Zhuang+2024} region are represented by the orange, cyan, and purple sectors, respectively. The origin of the coordinate is the Galactic center, and the distance to the Galactic center is 8.15 kpc. The best-fit spiral arm models from \citet{Reid+2019} are overlaid: Local (pink), Perseus (gold), and Sagittarius–Carina (purple).}
    \label{spatial_distribution}
\end{figure*}

Figure~\ref{spatial_distribution} illustrates the spatial distribution of our cataloged molecular clouds overlaid on the 3D dust map from \citet{Vergely+2022}. This background map, constructed via a hierarchical inversion of \textit{Gaia} DR2/EDR3 and 2MASS data, covers a volume of $6 \times 6 \times 0.8$~kpc$^3$ around the Sun. Our catalog shows excellent agreement with the large-scale dust structures revealed by \citet{Vergely+2022}, as well as with the distributions of molecular clouds from previous catalogs \citep{Yan+2021, Zucker+2020}. To further validate our results, we highlight three specific Galactic sectors where distance measurements exist from independent studies. In the Cygnus region (orange sector, $l \sim 72\degr$--$87\degr$), we identify two prominent cloud concentrations at $\sim 0.8$~kpc and $\sim 1.5$~kpc. These features correspond to the multi-layer extinction structures reported by \cite{Zhang+2024, Zhang+2025}, recovering approximately 70\% of the previously identified clouds. Similarly, in the Aquila Rift ($l \sim 20\degr$--$50\degr$, cyan sector) and the Monoceros OB1 region ($l \sim 198\degr$--$206\degr$, purple sector), we recover about 60\% and 50\% of the clouds reported by \cite{Su+2020} and \cite{Zhuang+2024}, respectively. The consistency between our distance estimates and these independent datasets supports both the accuracy and the high completeness of our catalog (see Table~\ref{Catalog}).

\begin{figure*}[!htbp]
    \centering
	\begin{minipage}{0.49\linewidth} 
        \centering
        \includegraphics[width=\linewidth]{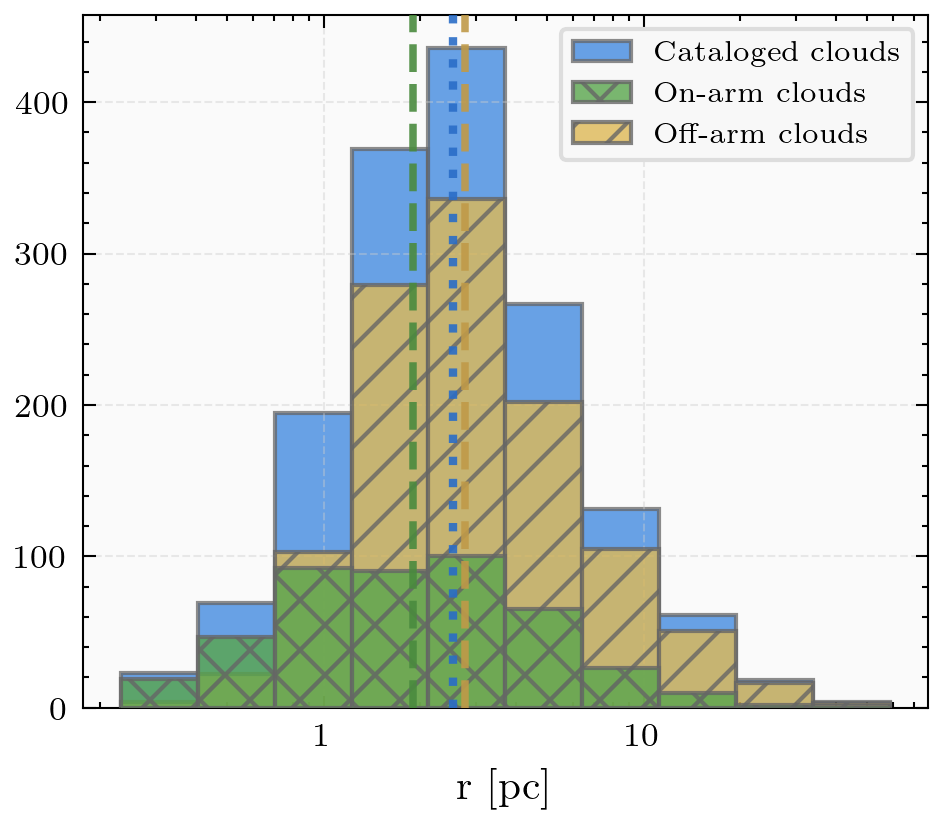}
        \label{Hist_rcpc}
    \end{minipage}
    \hfill 
    \begin{minipage}{0.49\linewidth} 
        \centering
        \includegraphics[width=\linewidth]{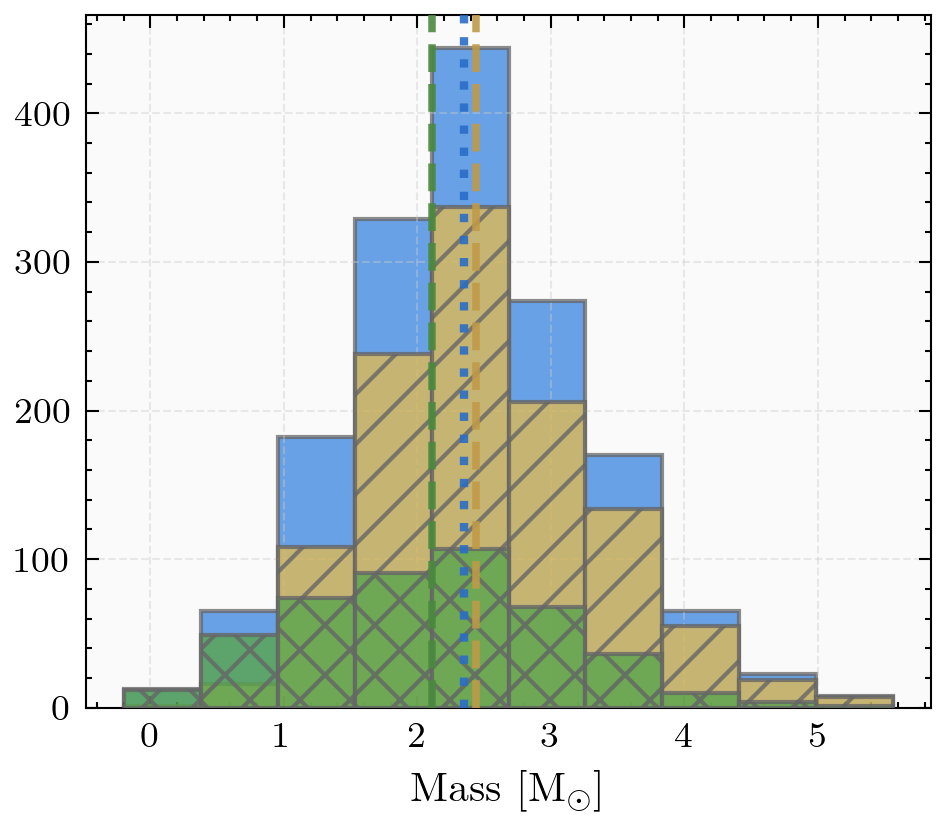}
        \label{Hist_mass}
    \end{minipage}
    \caption{The histogram distributions of the physical properties, including linear radii (left) and masses (right) of our cataloged molecular clouds. The colored vertical dashed lines are median values.}
    \label{Hist_rcpc_mass}
\end{figure*}

We further classify the sample based on the spiral arm model of \cite{Reid+2019}, dividing the clouds into those projected within the spiral arm width (``on-arm'') and those located in the inter-arm regions (``off-arm''). As shown in Figure~\ref{spatial_distribution}, the sample is numerically dominated by off-arm clouds. We present the distributions of linear radii ($r$) and masses for the full catalog, as well as for the on-arm and off-arm subsets, in Figure~\ref{Hist_rcpc_mass}. The full sample spans radii from $\sim 0.2$ to 58.7~pc (median 2.5~pc) and masses from $\sim 0.6$ to $3.6 \times 10^5~M_\odot$ (median $2.2 \times 10^2~M_\odot$). Interestingly, on-arm clouds appear systematically smaller and less massive (median $r=1.9$~pc, $M=1.3 \times 10^2~M_\odot$) compared to the off-arm population (median $r=2.7$~pc, $M=2.8 \times 10^2~M_\odot$). This trend may be partially introduced by the DBSCAN algorithm used for cloud identification; while effective at detecting weak emission, DBSCAN tends to fragment large, structured molecular complexes into smaller components, potentially biasing the statistics in dense arm regions.

\section{Summary}  
In this paper, we present a comprehensive catalog of 1,573 molecular clouds from the full Phase I data cubes of the MWISP survey, spanning the Galactic longitude range $l=9.75\degr-229.75\degr$ and latitude $|b| \leq 5.25\degr$. We adopted three methods to determine molecular cloud distances by combining $^{12}$CO integrated intensity maps with 3D extinction maps. Among the 10,929 MCs identified from the MWISP Phase I CO survey, we have measured distances for 1,573 clouds. These distances range from $\sim$150 to 3000 pc, and 90\% of them are determined for the first time. The typical statistical uncertainty of the distance is $\sim$ 20$\%$, and the systematic uncertainty is $\sim$ 10$\%$. We have also measured the physical properties of clouds, such as linear radius and mass. This catalog of molecular clouds with distances provides a critical foundation for testing classical empirical relations of molecular cloud and for investigating correlations with star formation activities across diverse Galactic environments.

\begin{acknowledgments}
This work is supported by the National Key R\&D Program of China (grant No. 2023YFA1608000). This research made use of the data from the Milky Way Imaging Scroll Painting (MWISP) project, which is a multi-line survey in $^{12}$CO/$^{13}$CO/C$^{18}$O along the northern galactic plane with PMO-13.7m telescope. We are grateful to all the members of the MWISP working group, particularly the staff members at the PMO-13.7 m telescope, for their long-term support. MWISP is sponsored by the National Key R\&D Program of China with grants 2023YFA1608000 and 2017YFA0402701, and the CAS Key Research Program of Frontier Sciences with grant QYZDJ-SSW-SLH047. ZC acknowledges the Natural Science Foundation of Jiangsu Province (grants No. BK20231509). M.Z. acknowledges the support of NSFC grant 12473026. Y.S. acknowledges funding from the Basic Research Program of Jiangsu (BK20252109). This work has made use of data from the European Space Agency (ESA) mission Gaia (\url{https://www.cosmos.esa.int/gaia}), processed by the Gaia Data Processing and Analysis Consortium (DPAC, \url{https://www.cosmos.esa.int/web/gaia/dpac/consortium}). Funding for the DPAC has been provided by national institutions, in particular the institutions participating in the Gaia Multilateral Agreement.
\end{acknowledgments}





%
\facilities{PMO 13.7 m, Science Data Bank (ScienceDB)}

\software{astropy \citep{Astropy+2022}, Numpy \citep{Harris+2020}, Matplotlib \citep{Hunter+2007}
          }




\bibliography{sample701}{}
\bibliographystyle{aasjournalv7}



\end{document}